\pdfoutput=1
\documentclass{aastex631}
\usepackage{amsmath}

\graphicspath{{./}}

\submitjournal{ApJ}

\shorttitle{Water Megamasers of NGC~1068}
\shortauthors{Gallimore \& Impellizzeri}

\newcommand{\water}{\hbox{H$_2$O}}

\newcommand{\E}[1]{\hbox{$10^{#1}$}}
\newcommand\x      {\hbox{$\times$}}
\newcommand\mic    {\hbox{$\mu$m}}
\newcommand\Mo     {\hbox{$M_{\sun}$}}
\newcommand\kms    {\hbox{km\,s$^{-1}$}}

\newcommand\mone   {\hbox{$^{-1}$}}
\newcommand\pv     {\hbox{$p$--$v$}}
\newcommand\vrot   {\hbox{$v_{\mathrm{rot}}$}}
\newcommand\vsys   {\hbox{$v_{\mathrm{sys}}$}}

\newcommand\ecc   {\hbox{$e$}}

\begin{document}


\title{High Sensitivity Observations of the \water{} Megamasers of NGC~1068: Precise Astrometry and Detailed Kinematics} 



\correspondingauthor{Jack F. Gallimore}
\email{jgallimo@bucknell.edu}

\author[0000-0002-6972-2760]{Jack F. Gallimore}
\affil{Department of Physics and Astronomy, 
Bucknell University, 
Lewisburg, PA 17837}

\author[0000-0002-3443-2472]{C. M. Violette Impellizzeri}
\affil{Leiden Observatory, Leiden University, PO Box 9513, 2300 RA, Leiden, The Netherlands}

\begin{abstract}
We present High Sensitivity Array observations of the \water{} megamasers of NGC~1068. We obtain absolute astrometry with 0.3~mas precision that confirms the association of the disk masers with the nuclear radio continuum source S1. The new observations reveal two new blueshifted groups of disk masers. We also detect the 22~GHz continuum on short interferometric baselines. The position-velocity diagram of the disk masers shows a curve consistent with a nonaxisymmetric distribution of maser spots. This curve is probably the result of spiral arms with a constant pitch angle $\sim 5\degr{}$. The disk kinematics are consistent with Keplerian rotation and low turbulent speeds. The inferred central mass is $17\times 10^6$~\Mo{}. On the basis of disk stability arguments, the mass of the molecular disk is {$\approx 110 \times 10^3$~\Mo{}}. The disk masers further resolve into filamentary structures suggesting an ordered magnetic field threading the maser disk. {The magnetic field strengths must be $\ga 1.6$~mG to withstand turbulent motions in the partially ionized molecular gas.} We note apparent asymmetries in the molecular disk that might be explained by anisotropic heating by a misaligned inner accretion disk. The new observations also detect the fainter jet masers north of the disk masers. The distribution and kinematics of the jet masers are consistent with an expanding ring of molecular gas. 
\end{abstract}


\section{Introduction} \label{sec:intro}

Extragalactic \water{} megamasers are found to be associated with molecular accretion disks and nuclear outflows (\water{} megamasers are reviewed in \citealt{2005ARA&A..43..625L}). The most famous case is NGC~4258 \citep{1999Natur.400..539H}, in which \water{} masers trace Keplerian rotation in a sub-pc scale, warped disk, and new megamaser sources have been discovered and studied in recent years \citep{2011ApJ...727...20K}. \water\ megamaser disks are of broader astrophysical importance because they provide tight constraints on the centrally concentrated masses of galaxies, presumably supermassive black holes (e.g., \citealt{2017ApJ...834...52G}), and, with the measurement of centripetal accelerations, they afford direct, geometrical measurement of distances to galaxies (e.g., \citealt{2016ApJ...817..128G}).  

NGC~1068 is the archetypal hidden type 2 Seyfert galaxy \citep{1985ApJ...297..621A} and one of the first AGNs shown to harbor a circumnuclear \water{} megamaser disk \citep{1996ApJ...462..740G, 1996ApJ...472L..21G}. Two different sources of \water{} megamaser emission were identified in NGC~1068, one associated with the compact radio source ``C,'' which appears to mark the interaction between the radio jet and a molecular cloud; and the other with the nuclear radio source ``S1'' \citep{1996ApJ...462..740G,2001ApJ...556..694G}. In the commonly accepted interpretation, the S1 megamasers trace a nearly edge-on, geometrically thin annular disk surrounding the central engine. This interpretation is partly inspired by their resemblance to the \water\ megamasers of NGC~4258. Primarily, the S1 megamasers show the classic triply-peaked spectrum expected from an edge-on rotating disk (or annulus) of molecular gas \citep{1994ApJ...432L..35W,1996ApJ...462..740G, 2001ApJ...556..694G}, although the redshifted masers are consistently brighter in monitoring observations \citep{2001ApJ...556..694G}. For the purpose of discussion, we refer to the masers associated with radio component C as jet masers, and those with component S1 as disk masers.

Using the Very Long Baseline Array (VLBA) augmented by the phased Very Large Array (VLA), \cite{gg97} (GG97) presented the first VLBI observations of the disk masers that cover the full 800--1500~\kms{} velocity range of the maser spectrum. Discussed in further detail in Section~\ref{sec:data}, the disk maser spots roughly align along PA~$-50\degr$ and span about 1.75~pc.\footnote{The distance to NGC~1068 is $13.97\pm 2.1$~Mpc \citep{2021MNRAS.501.3621A}. For convenience and consistency with previous papers, we adopt the scale 1\arcsec = 70~pc, appropriate for a distance of 14.4~Mpc.} Based on a preliminary model for the maser kinematics, GG97 estimated the systemic recessional velocity\footnote{To avoid confusion between the disk-frame and sky-frame velocities, we denote motion relative to the observer as recessional velocity, and radial velocities refer to radial motions in the disk-frame.} of the maser disk, $V_{LSR} = 1119$~\kms{} (optical convention). The redshifted masers show decreasing recessional velocities with distance from the center of the maser spot distribution in the sky and outside the inferred inner radius of $R_{in} \approx 0.6$~pc. The fainter, blueshifted masers are more tightly grouped on the sky and therefore do not sample the velocity gradient as well as the redshifted masers. 
Recently, \cite{2022PASJ..tmp..119M} presented global VLBI observations made in February 2000. Their results largely confirm those of GG97, although they claim the detection of faint disk maser spots displaced northeast and southwest of the molecular disk, i.e., along the outflow axis.

The conventional interpretation for the position-velocity diagram has been that the high-velocity maser spots trace molecular clumps that are viewed along sight lines nearly tangential to their orbits. In other words, their observed recessional velocities trace the actual rotational velocities relatively unaffected by projection, so the declining velocities are a direct measure of the disk rotation curve. With this assumption, \cite{1996ApJ...472L..21G} reported a sub-Keplerian\footnote{Here, sub-Keplerian means having rotation curves falling more slowly than $r^{-0.5}$, which translates to rotational velocities greater than the expected Keplerian velocity with increasing radius. In other words, sub-Keplerian means a rotation curve that is flatter than Keplerian.} rotation curve, $v\propto r^{-0.31}$, where they assumed the major axis lies along position angle $-90\degr$. \cite{LB03}, analyzing the data of GG97, find that the rotation curve varies from $v\propto r^{-0.35}$ at the inner edge to $v\propto r^{-0.30}$ at the outer edge with the major axis assumed to lie along position angle $-45\degr$. Performing a similar analysis, \cite{2022PASJ..tmp..119M} argue for a flatter rotation curve, $v\propto r^{-0.24}$.  

\cite{Kumar99}, \cite{2002A&A...395L..21H}, and \cite{LB03} took the apparent sub-Keplerian rotation curve as evidence of disk self-gravity. Kumar argued that a standard $\alpha$-disk analysis requires the disk to have a mass of nearly $10^8\ \mbox{M}_\odot$, greatly exceeding the inferred black hole mass. \cite{2002A&A...395L..21H} inverted the rotation curve to infer the disk mass; in this analysis, the disk mass is about 75\% of the central black hole mass. \cite{LB03} presented a model of a self-gravitating disk that self-regulates against the Jeans instability to explain the sub-Keplerian rotation curve. In their model, the central black hole mass and the accretion disk mass roughly balance with $M_{bh} \approx M_{disk} \approx 8\times 10^6\ \mbox{M}_\odot$.

Unlike the rotation curve model for the disk masers, recent ALMA observations of HCN kinematics $(J = 3\rightarrow 2)$ show a tangential velocity curve that is consistent with Keplerian (counter) rotation on the radial scales 1.4 to 7~pc from the kinematic center, just outside the maser disk \citep{2019ApJ...884L..28I}. Further to the point, the HCN tangential velocity curve is not consistent with the proposed flatter rotation curve of the \water{} maser disk; the extrapolated rotation speeds exceed the HCN tangential velocities by up to 60\%. To reconcile the difference, it seems that one or the other, the redshifted \water{} masers or the HCN tangential velocity curve, does not directly reflect the rotation curve. 

We have observed the \water{} megamasers of NGC~1068 using the High Sensitivity Array, a combination of the Very Long Baseline Array (VLBA), the phased Karl G. Jansky Very Large Array (VLA), and the Robert C. Byrd Green Bank Telescope (GBT). Our main goal was to try to recover fainter disk maser spots in hopes of better constraining the subparsec rotation curve, but we were also able to determine absolute astrometry of the \water{} maser spots and recover low-surface-brightness radio continuum emission. We also detected jet masers associated with the radio continuum component C.  Section~\ref{sec:data} describes the observations and astrometric analysis. Section~\ref{sec:results} presents the primary results, including recovery of the 22~GHz continuum and the distribution and kinematics of the \water{} maser spots. We consider three different kinematic models to explain the peculiarities of the position-velocity diagram; the details are provided in Section~\ref{sec:kinematicmodels}. The radically improved astrometry of the maser spot positions relative to the nuclear radio continuum source affects the interpretation of infrared observations of the nuclear obscuring region; the implications of the improved astrometry are discussed in Sec.~\ref{sec:infrared}. In Section~\ref{sec:magneticFields}, we consider how magnetic fields may affect the kinematics and structures of the \water{} maser disk. The kinematics of the jet masers is discussed in Section~\ref{sec:jetmasers}. The primary conclusions are summarized in Section~\ref{sec:conclusions}. 

\section{Observations, Calibration, and Data Reduction} \label{sec:data}

We observed NGC~1068 with the HSA on 8--9 February 2020 (BG262D) and again on 21--22 March 2020 (BG262J). Each observation, including the calibrators and overhead, spanned 6 hours. For both observations, the receivers were tuned to the \water{} maser transition, $\nu_0 = 22235.080$~MHz, and adjusted to $V_R = 1150$~\kms{} recessional velocity (LSRK, optical convention). The HSA maintains fixed frequencies in the topocentric frame during the course of the observation, so the receivers were tuned to this recessional velocity at the midpoint of the observation. For reference, the systemic velocity of the host galaxy is $\vsys(\mbox{host}) = 1132\pm 5$~\kms{}  \citep{2003A&A...412...57P}, and the systemic velocity of the molecular disk of the pc scale surrounding the maser ring is $\vsys(\mbox{host})\simeq 1133\pm 3$~\kms{} \citep{2019ApJ...884L..28I}. We used a single IF band with bandwidth $\Delta \nu = 64$~MHz, corresponding to a radial velocity range $\Delta V_R \approx 870$~\kms. The channel widths are determined at the time of correlation. For BG262D, the IF was divided into 1024 channels (0.85~\kms{} channels), and for BG262J, 2048 channels (0.42~\kms{} channels). 

Unfortunately, a software error affected the BG262D observations. Roughly one-third of the observing time with the phased VLA was lost, and only one polarization of the remaining observations was recovered. We also found that the data that included the Hancock antenna were not usable. As a result, we focus our analysis on the results of BG262J, although we also processed the BG262D data and used them to check astrometric uncertainties.

The observations consisted of scans of fringe calibrators (3C454.3 and 3C84, which are also used as bandpass calibrators), and alternating scans of NGC~1068 and the phase reference calibrator J0239$-02$, located 2\fdg7 away. The cadence was typically 9 minutes at the source and 3 minutes at the phase reference. However, this cadence was periodically interrupted to allow calibration of the GBT (roughly once per hour), phasing of the VLA (up to four times per hour), and pointing calibration of the VLA (twice per hour). While the VLA or GBT performed calibrations, the remaining antennas continued to observe the source and phase reference. Phased VLA observations were also interrupted for a 5~minute scan of 3C48 to establish the flux scale.

Data reduction and calibration followed standard procedures in AIPS \citep{1985daa..conf..195W,1990apaa.conf..125G,2003ASSL..285..109G}. First, the data were corrected for terrestrial effects, including ionospheric Faraday rotation, dispersive delay, and updated Earth Orientation Parameters. The data were then corrected for digital sampling artifacts, and a bandpass calibration was generated based on the observations of 3C454.3 and 3C84. After these frequency-dependent calibrations were applied, the observations were shifted in frequency to the LSRK reference frame. Next, we performed fringe fitting on J0239$-$02 to determine the initial calibration of the phase rates and delays. After this calibration, the masers were apparent on the cross-power spectrum (a spectrum derived from a time average of visibilities), and the brightest masers were found at recessional velocities between $V_R = 1409.5$~\kms{} and 1415.4~\kms. We determined a phase-only self-calibration based on the brightest maser spots identified in these channels and applied the correction back to the phase reference, J0239$-$02. 

\subsection{Astrometric Calibration}\label{sec:astrometry}

\begin{figure*}[tbh]
  \centering
  \plottwo{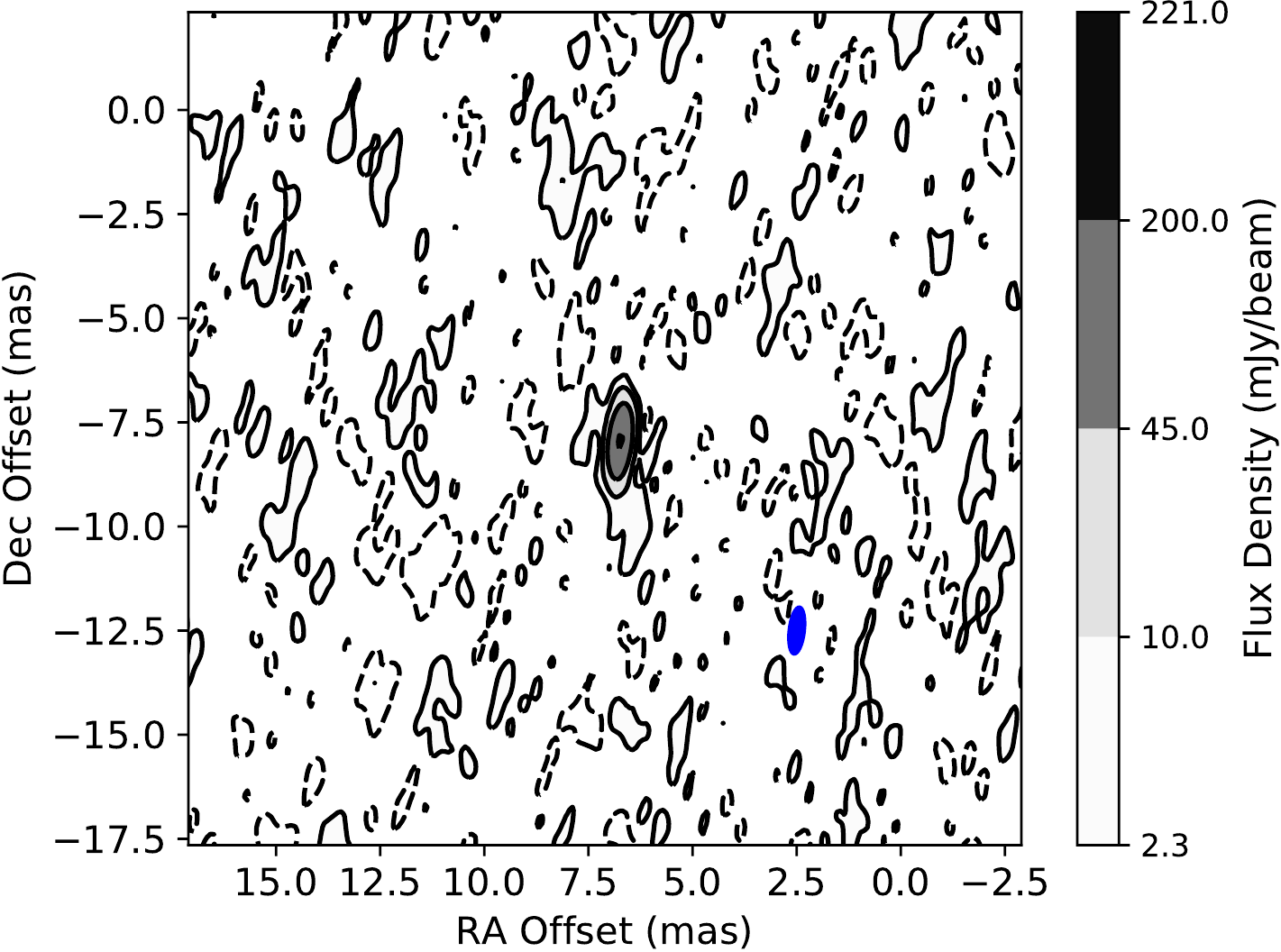}{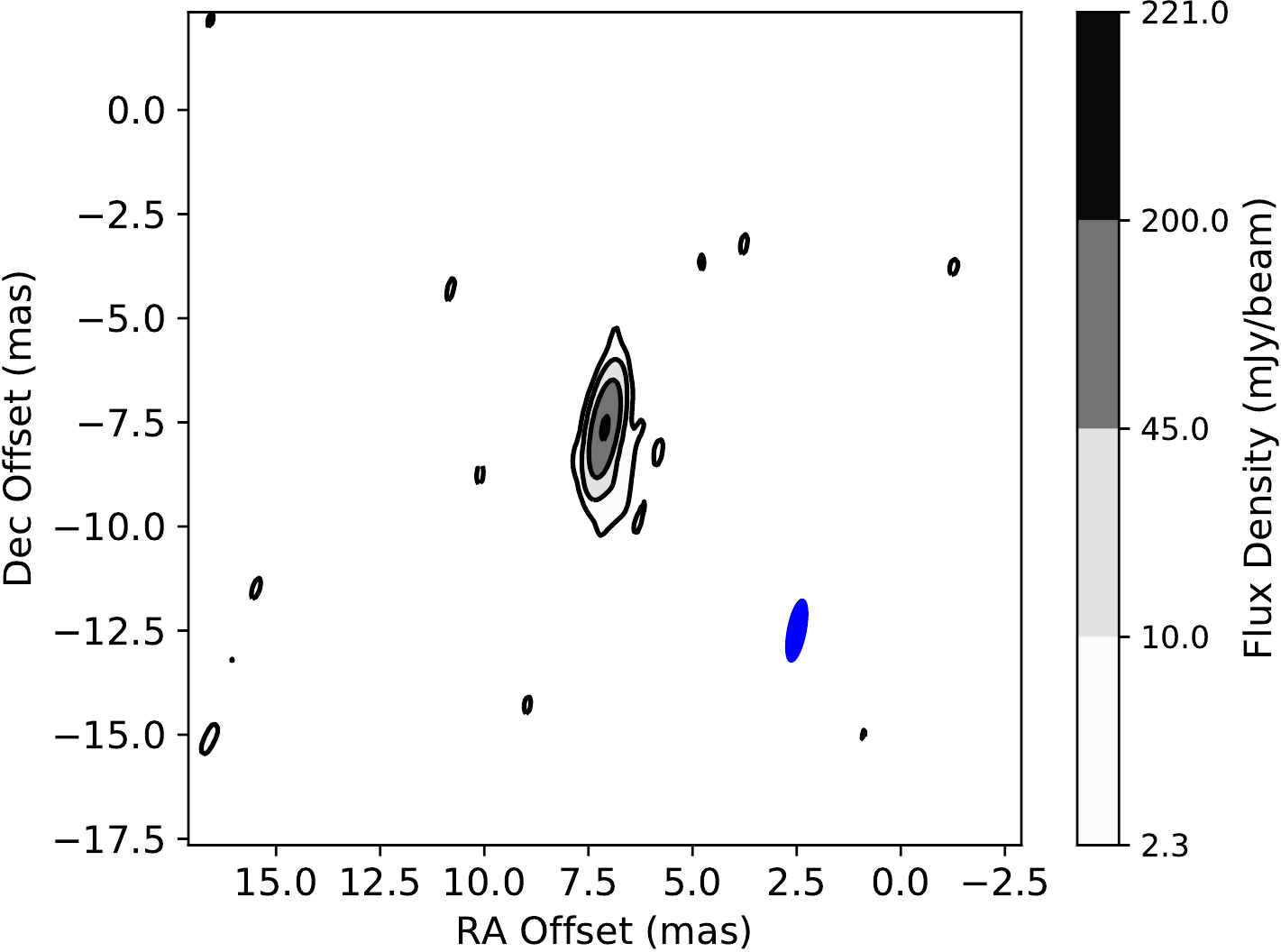}
\caption{Naturally-weighted 22~GHz continuum images of the phase reference source, J0239$-$02. The image derived from the BG262D observations is shown on the left, and BG262J is shown on the right. The coordinates are offsets relative to the VLBA Calibrator Catalog position, $\alpha$(J2000) = 02h~39m~45\fs472272, $\delta$(J2000) = $-$02\degr~34\arcmin~40\fdg99146. The displacement from the nominal position results from initial phase calibration based on the brightest masers of NGC~1068, which we use to calibrate the absolute astrometry of the maser spots. As depicted in the colorbar, the contours are $\pm 2.3$, 10, 45, and 200~mJy~beam\mone. The BG262D image is noisier primarily due to a loss of time on the phased VLA and the Hancock antenna. The restoring beams are shown as filled blue ellipses.\label{fig:phaseReference}}
\end{figure*}

The data were phase-referenced to the brightest maser spots, and, prior to this work, the absolute astrometry of these spots was accurate to about 5~mas \citep{2001ApJ...556..694G}. This phase calibration introduces an offset to the position of the phase reference relative to the coordinates of the pointing center. The inverse of this offset corrects the positions of the maser spots with respect to the astrometric frame. To this end, we made images of the phase reference source and measured the sky offset from the pointing center. The images of J0239$-$02 are presented in Figure~\ref{fig:phaseReference}. For reference, the astrometric precision of J0239$-$02 is reported as 0.03~mas in the \href{http://www.vlba.nrao.edu/astro/calib/}{VLBA Calibrator Catalog}. In the BG262J observations, J0239$-$02 was offset by $\Delta(\mbox{RA})= +7.03$~mas, $\Delta(\mbox{Dec}) = -7.57$~mas.  We repeated the measurement for the BG262D observations, and we measured sky offsets $\Delta(\mbox{RA}) = +6.73$~mas, $\Delta(\mbox{Dec}) = -7.88$~mas.  

Due to data loss that affected the BG262D observations, we adopted the corrections derived from the BG262J observations but used the BG262D offsets to estimate the systematic uncertainties of the phase calibration transfer between the source and phase reference. The estimated systematic uncertainties of this new absolute astrometric calibration are 0.4~mas in RA and 0.3~mas in Dec. For astrometric experiments, the characteristic uncertainty is $\sim 0.05$~mas, but poor weather and data loss can degrade the uncertainty to $\sim 0.2$~mas \citep{2006A&A...452.1099P}. Furthermore, phase referencing was not a primary goal of the BG262 observations, and the observing cadence was not designed for the highest possible astrometric accuracy. Nevertheless, the astrometric uncertainties are an order of magnitude improvement compared to previous work. The uncertainties are small compared to the distribution of masers, approximately 30 mas in extent, and the size of the continuum source, roughly 10~mas. The centroid position of the 22~GHz continuum source (Section~\ref{sec:continuumReduction}) agrees to within 0.4~mas of the 5~GHz VLBA continuum position, comparable to the statistical uncertainty of the centroid position \citep{2004ApJ...613..794G}.  For purposes of comparing the continuum morphology and maser spot distribution, the new, 22~GHz continuum image was produced using the same data cube as the maser spot measurements, and signal-to-noise ratio primarily limits the relative astrometric measurements between the continuum and maser spots rather than systematic calibration effects. 

For consistency with previous publications, the absolute positions of the NGC~1068 data have been referenced as offsets relative to the VLBA 5~GHz centroid of radio continuum source S1, RA(J2000) = 02h~42m~40\fs70905, Dec(J2000) = $-$00\degr~00\arcmin~47\farcs945 \citep{2004ApJ...613..794G}.

\subsection{Maser Spot Measurements}\label{sec:spotMeasurements}

Final processing and imaging were performed using DIFMAP \citep{1997ASPC..125...77S}. Additional self-calibration of the maser data cubes was performed using the bright maser spot at $V_R = 1414.3$~\kms{} as a reference. We produced naturally weighted images of each spectral channel. The restoring beam is $1.12 \times 0.36$~mas, PA $-9\fdg9$. The characteristic background rms is $1.3$~mJy~beam\mone{} in a single channel; masers as faint as about 5~mJy~beam\mone{} could be detected. Based on the background rms of the channel with the brightest maser emission, the dynamic range is about 60. As a result, in channels containing the brightest masers, the limiting point source sensitivity is about 20~mJy~beam\mone.

\begin{figure*}[tbh]
  \centering
\plotone{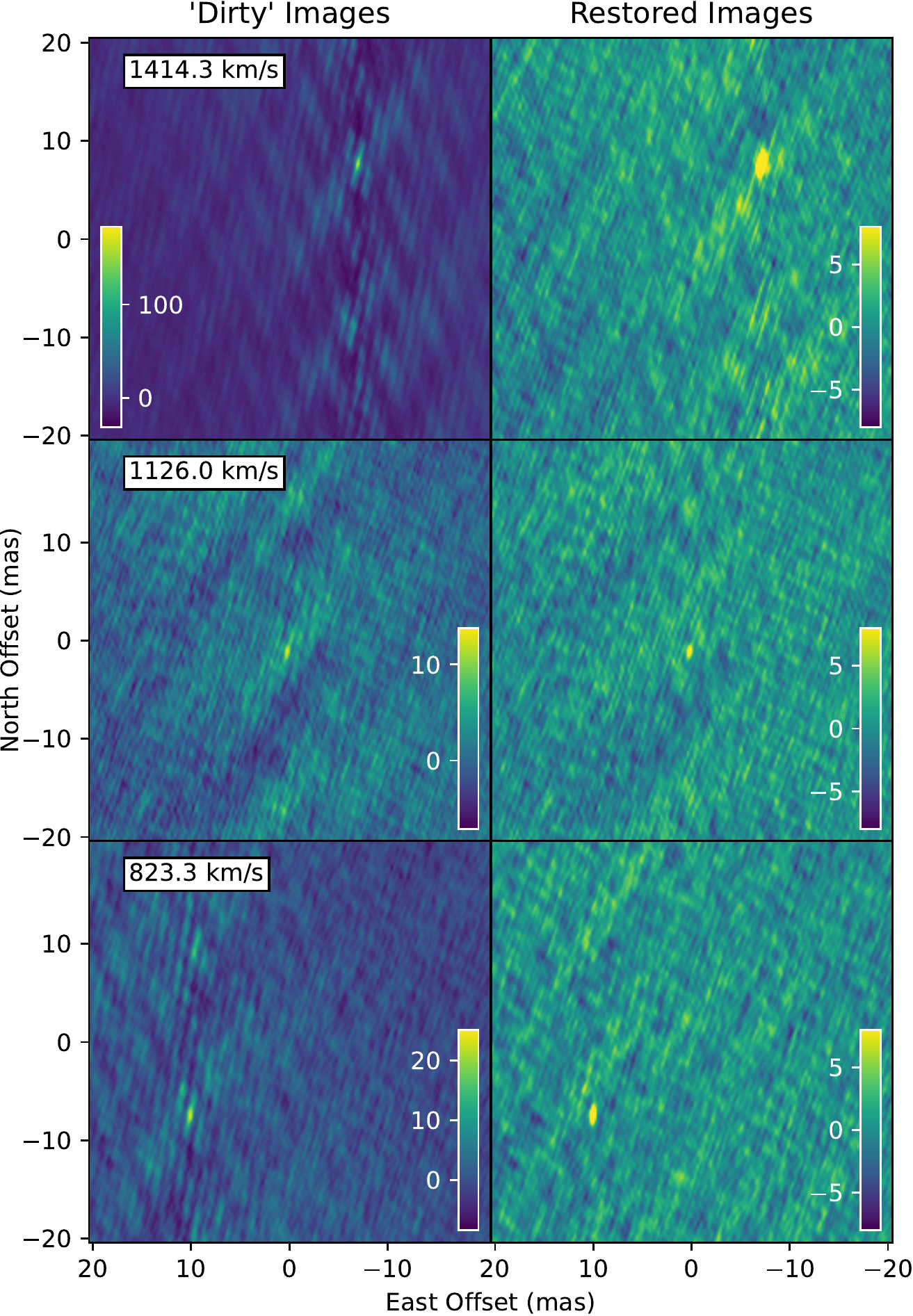}
\caption{Naturally-weighted channel maps of the disk masers. Each row is a separate channel; the recessional velocities of each channel are labeled in the left panel. The first column is the ``dirty'' map created by Fourier inversion, and the second column is the restored image produced by fitting Gaussian source models. The image stretch in mJy~beam\mone{} is shown as a colorbar inset in each figure. For the dirty maps, the stretch covers the entire range of flux densities in the map; for the restored images, the stretch is truncated at $\pm 8$~mJy~beam\mone{}, or roughly $\pm 5\sigma$, to illustrate residual sidelobe artifacts in the restored image. \explain{We added this figure to illustrate the image restoration process and highlight artifacts that remain after restoration.} \label{fig:channelmaps}}
\end{figure*}

We identified the channels that show clear evidence of compact maser emission within a region $82 \times 82$~mas centered on the position of the continuum source S1. Similarly, we searched for maser spots in the region of the continuum source C, located roughly 60~mas east and 290 mas north of S1. The position and brightness of each detected maser spot was determined by a two-dimensional Gaussian fit to the visibility data (DIFMAP task {\tt modelfit}). {To illustrate, Figure~\ref{fig:channelmaps} shows the ``dirty'' and restored channel maps of three disk maser sources.}  In some channels, two or more maser spots were present, and so multiple Gaussians were included in the model. {We specifically avoided  adding model components to regions roughly 20~mas north or south of a bright maser spot to avoid sidelobe artifacts.} The best-fit photocenters were then shifted to the absolute coordinate frame. {The results are provided in Tables~\ref{tab:s1spots} and \ref{tab:ccspots}.}

\begin{deluxetable}{cDD@{$\pm$}DD@{$\pm$}DD@{$\pm$}Dl}
\tablecaption{Disk Maser Spot Positions and Velocities}\label{tab:s1spots}
\tablehead{Channel & \multicolumn2c{$V$(LSRK)} & \multicolumn4c{Flux Density} & 
\multicolumn4c{East Offset} & \multicolumn4c{North Offset} & Group\\
& \multicolumn2c{(\kms{})} & \multicolumn4c{(mJy)} & 
\multicolumn4c{(mas)} & \multicolumn4c{(mas)} & } 
\decimals
\startdata
 479 & 1381.63 & 22.8 & 1.8  & -8.853 & 0.015 & 13.379 & 0.064 & R1 \\
 480 & 1381.20 & 29.6 & 2.1  & -8.906 & 0.011 & 13.560 & 0.042 & R1 \\ 
 481 & 1380.78 & 22.8 & 1.8  & -8.873 & 0.015 & 13.323 & 0.068 & R1 \\ 
 482 & 1380.36 & 13.5 & 1.6  & -8.871 & 0.030 & 13.537 & 0.130 & R1 \\ 
 483 & 1379.93 &  7.8 & 1.0  & -8.900 & 0.021 & 13.629 & 0.060 & R1 \\ 
 437 & 1399.46 & 21.5 & 2.3  & -9.193 & 0.019 & 10.523 & 0.052 & R2a \\ 
\enddata
\tablecomments{Table~\ref{tab:s1spots} is published in its entirety in machine-readable format. A formatted sample of the data is provided here. The offset positions are relative to the VLBA 5~GHz continuum position of S1, RA(J2000) = 02h~42m~40\fs70905, Dec(J2000) = $-$00\degr~00\arcmin~47\farcs945.}
\end{deluxetable}

\begin{deluxetable}{cDD@{$\pm$}DD@{$\pm$}DD@{$\pm$}D}
\tablecaption{Jet Maser Spot Positions and Velocities}\label{tab:ccspots}
\tablehead{Channel & \multicolumn2c{$V$(LSRK)} & \multicolumn4c{Flux Density} & 
\multicolumn4c{East Offset} & \multicolumn4c{North Offset}\\
& \multicolumn2c{(\kms{})} & \multicolumn4c{(mJy)} & 
\multicolumn4c{(mas)} & \multicolumn4c{(mas)}  } 
\decimals
\startdata
1321 & 1024.44  & 7.2 & 0.8 & 65.86 & 0.02 & 280.99 & 0.08 \\ 
1325 & 1022.74  & 5.9 & 0.8 & 65.88 & 0.03 & 281.20 & 0.10 \\ 
1327 & 1021.90  & 6.8 & 0.7 & 65.86 & 0.03 & 281.03 & 0.09 \\ 
1328 & 1021.47  & 5.8 & 0.7 & 55.82 & 0.03 & 281.00 & 0.10 \\ 
1329 & 1021.05  & 5.6 & 0.7 & 65.81 & 0.03 & 281.20 & 0.10 \\ 
1330 & 1020.62  & 6.7 & 0.8 & 65.84 & 0.03 & 281.11 & 0.09 \\ 
\enddata
\tablecomments{Table~\ref{tab:ccspots} is published in its entirety in machine-readable format. A formatted sample of the data is provided here. The offset positions are relative to the VLBA 5~GHz continuum position of S1, RA(J2000) = 02h~42m~40\fs70905, Dec(J2000) = $-$00\degr~00\arcmin~47\farcs945.}
\end{deluxetable}

\begin{figure*}[tbh]
  \centering
\plotone{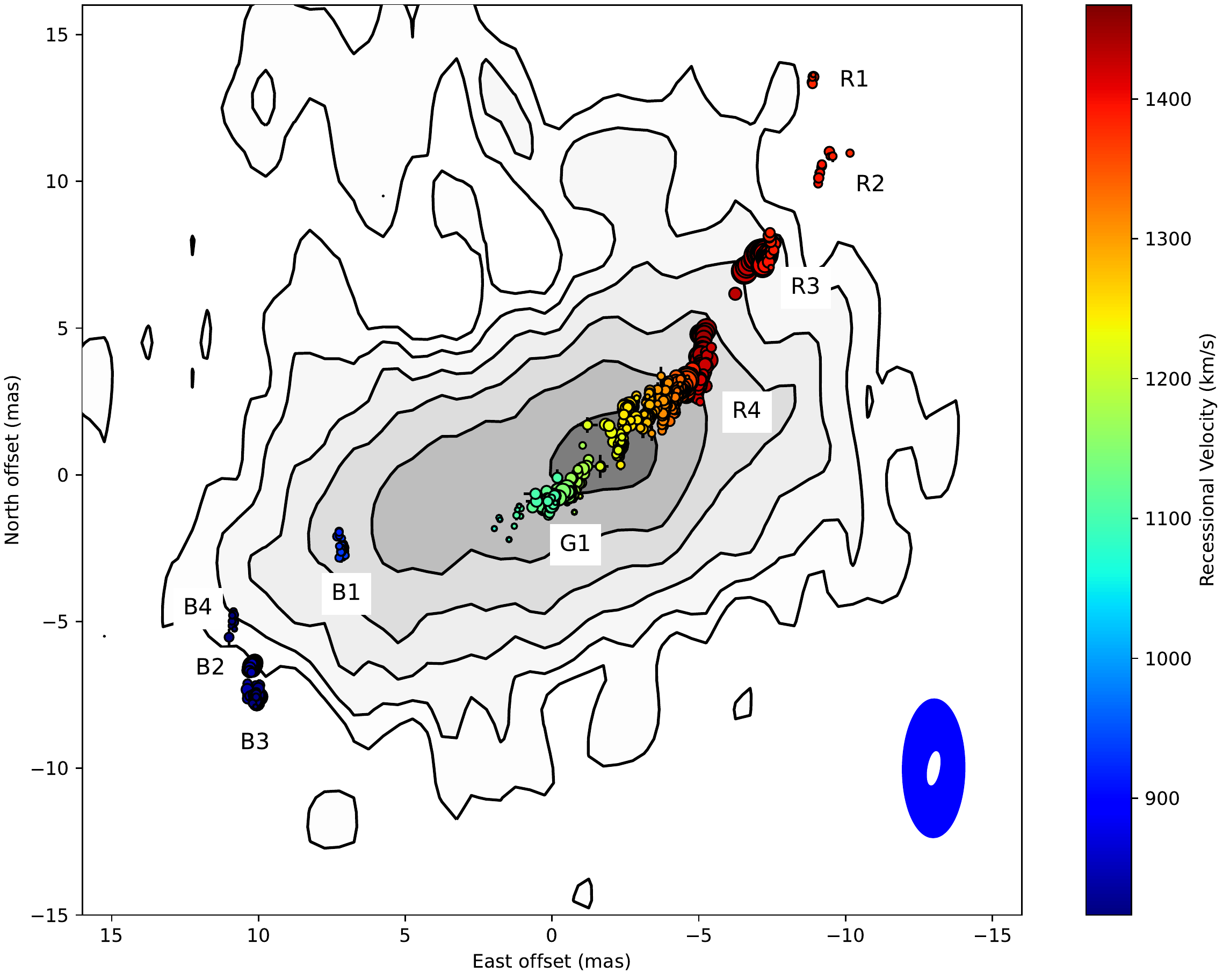}
\caption{Sky map of the nuclear \water{} maser spots. The locations of the spots are plotted as color-filled circles. The size of the symbols scales with the flux density of the maser spot, and the symbols are color-coded by recessional velocity as shown in the colorbar. The spots are plotted atop the 5~GHz continuum contours from \cite{2004ApJ...613..794G}.  The contour levels are $\pm 0.11$, 0.16, 0.24, 0.35, 0.51, 0.75~mJy~beam\mone{}. The 5~GHz beam is shown as the blue ellipse on the lower right; the 22~GHz beam is the white ellipse. The sky coordinates are offsets relative to the VLBA 5~GHz continuum position of S1, RA(J2000) = 02h~42m~40\fs70905, Dec(J2000) = $-$00\degr~00\arcmin~47\farcs945. \label{fig:s1-skyplot}}
\end{figure*}

\begin{figure*}[tbh]
  \centering
\plotone{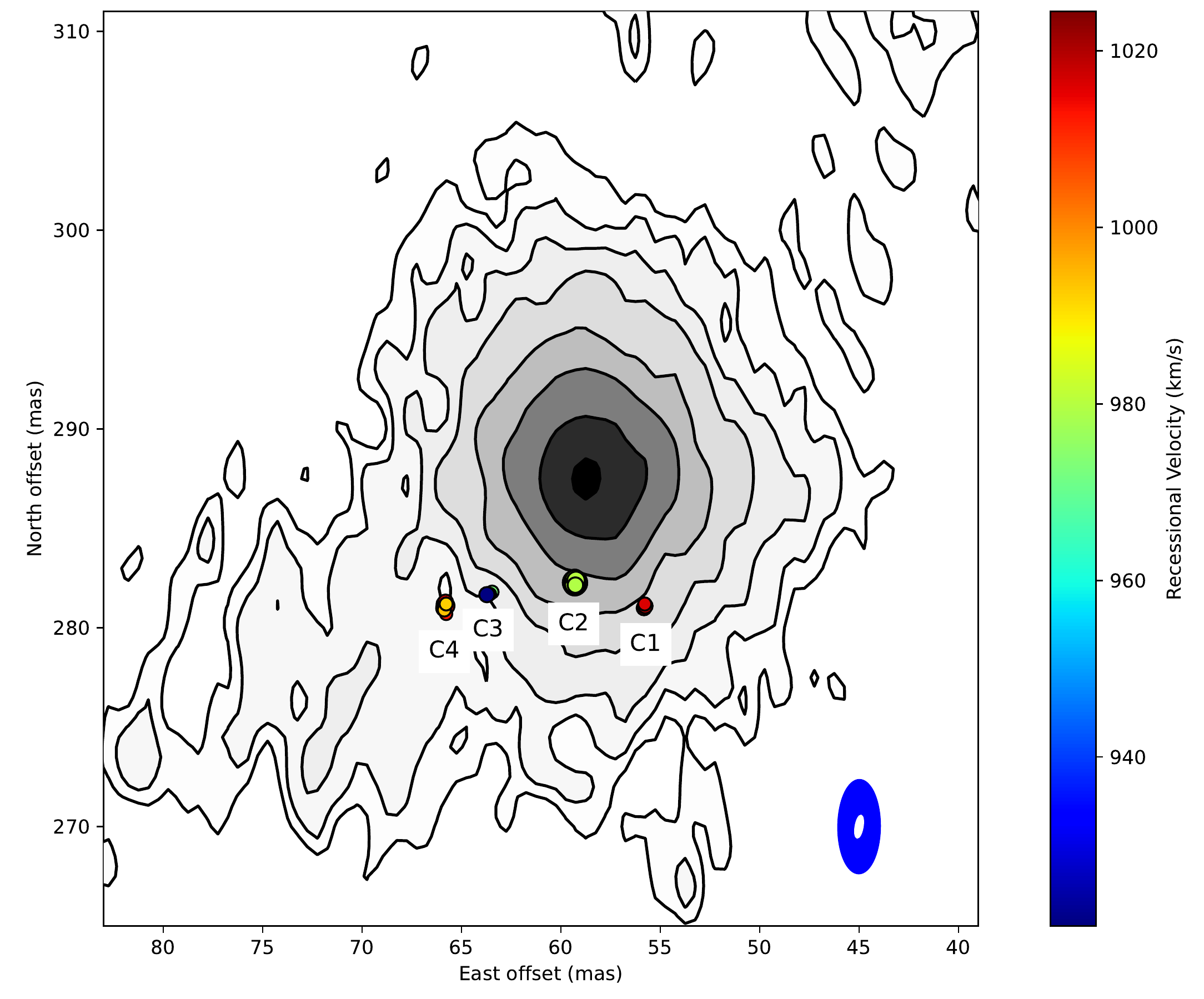}
\caption{Sky map of the jet \water{} maser spots. The locations of the spots are plotted as color-filled circles. The size of the symbols scales with the flux density of the maser spot, and the symbols are color-coded by recessional velocity as shown in the colorbar. The spots are plotted atop the 5~GHz continuum contours from \cite{2004ApJ...613..794G}. The contour levels are $\pm 0.11$, 0.16, 0.24, 0.35, 0.51, 0.75, 1.10, and 1.50~mJy~beam\mone{}. The 5~GHz beam is shown as the blue ellipse on the lower right; the 22~GHz beam is the white ellipse. The sky coordinates are offsets relative to the VLBA 5~GHz continuum position of S1, RA(J2000) = 02h~42m~40\fs70905, Dec(J2000) = $-$00\degr~00\arcmin~47\farcs945. \label{fig:c-skyplot}}
\end{figure*}

The sky map of the disk maser spots is shown in Figure~\ref{fig:s1-skyplot}, and the jet maser spots are shown in Figure~\ref{fig:c-skyplot}. The disk maser spots follow the pattern found in GG97, with the redshifted masers located northwest of the systemic masers and the blueshifted masers to the southeast. Our astrometric data confirm the results of \cite{2001ApJ...556..694G}: the S1 masers extend across the resolved radio continuum source, and the near-systemic masers pass within 1~mas of the radio continuum centroid. {Even though the new HSA observations are about 2.5 times more sensitive,} we do not find evidence for disk masers along the jet / outflow axis as reported by \cite{2022PASJ..tmp..119M}. {However, those reported outflow masers would appear at positions adversely affected by sidelobe residuals (cf. Figure~\ref{fig:channelmaps}) and so were avoided in our analysis.}

The disk maser spots form distinct groups in space and radial velocity. For the purposes of discussion, we identified nine groups, R1--4 (redshifted maser groups), G1 (near systemic) and B1-4 (blueshifted); the group labels appear as annotations in Figure~\ref{fig:s1-skyplot}. Groups B1 and B4 were not detected in GG97. The masers with the highest recessional velocities are found in group R4, and the lowest velocity masers are located in groups B2 and B3. There is a nearly continuous distribution of spots between the velocity extremes of R4 and G1; we distinguish these groups only by a small gap in the sky distribution and recessional velocities. 

The jet masers are divided into four distinct spot groups oriented roughly east-west and spanning $\sim 10$~mas; see Fig.~\ref{fig:c-skyplot}. We label the spot groups C1 (west) through C4 (east). Interestingly, the jet masers appear to be significantly offset south of the local continuum peak. More specifically, the C2 maser group is located about 5.4~mas south of component C. In contrast to the nuclear groups, the jet maser groups show no significant filamentary substructure on the sky. Furthermore, the east-west distribution of the groups is very different from the pattern reported by \cite{2022PASJ..tmp..119M}; in their analysis, the jet masers are distributed in a ring with diameter $\sim 20$~mas.

\section{Results and Analysis} \label{sec:results}

\subsection{22~GHz Continuum}\label{sec:continuumReduction}

\begin{figure*}[tbh]
  \centering
\plotone{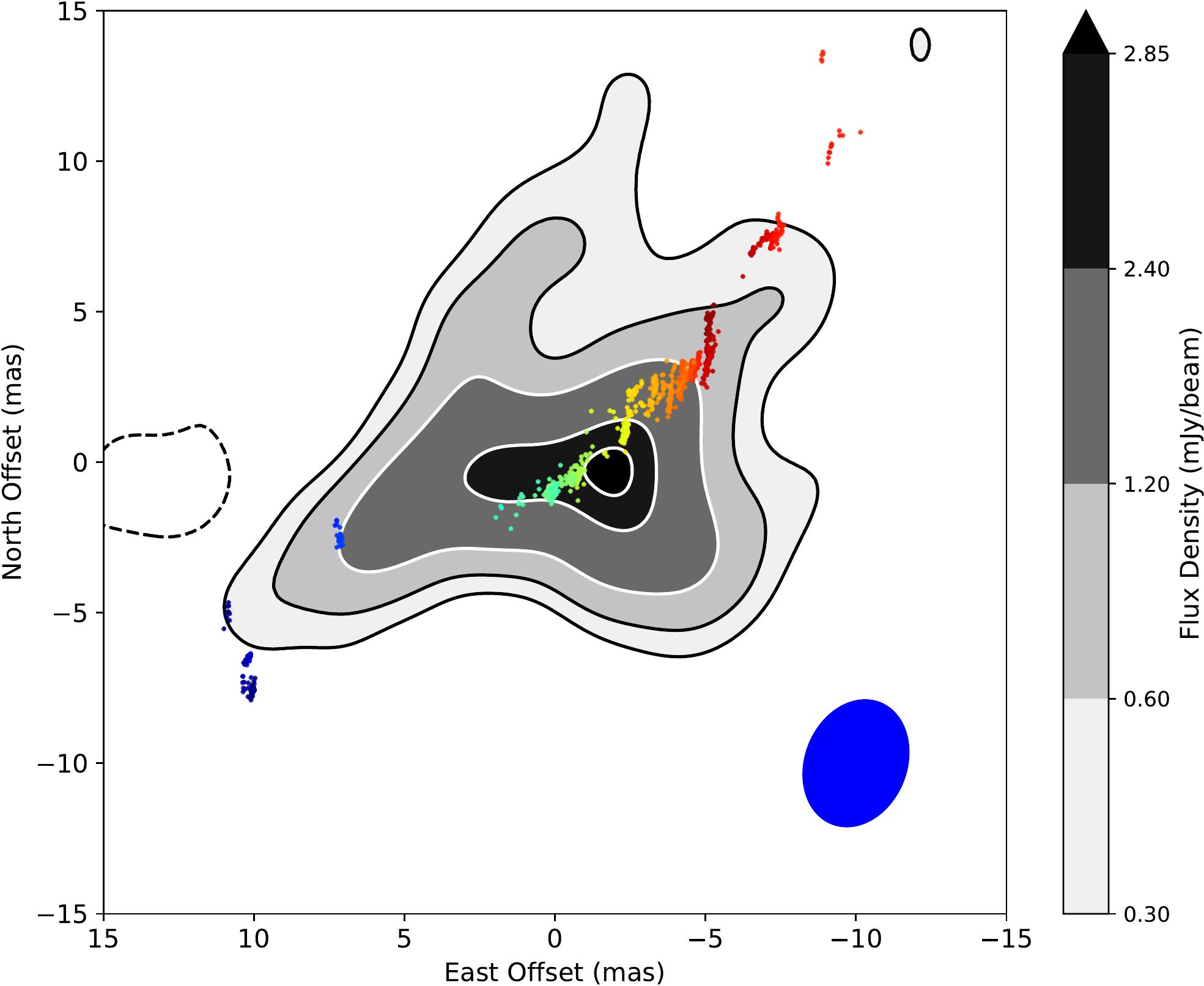}
\caption{Tapered 22~GHz continuum image of the nuclear radio source S1, recovered from line-free channels. As depicted in the scale bar on the right, the contours are $\pm 0.3$, 0.6, 1.2, 2.4, and 2.85~mJy~beam\mone. The maser spots are plotted as colored dots. The sky coordinates are offsets relative to the VLBA 5~GHz continuum position of S1, RA(J2000) = 02h~42m~40\fs70905, Dec(J2000) = $-$00\degr~00\arcmin~47\farcs945. The restoring beam for the continuum image is shown as the filled blue ellipse in the lower right.\label{fig:taperedContinuum}}
\end{figure*}

The surface brightness of the 22~GHz continuum emission is too low to detect on spectral line channels. However, weak continuum emission appears on short baselines after averaging line-free channels. To produce a continuum image, we averaged the line-free channels (16.3~MHz total bandwidth) and applied a Gaussian taper with 50\% weight at 30~M$\lambda$ during Fourier inversion. The data were naturally weighted during inversion, and the resulting image was deconvolved using the {\tt clean} task in DIFMAP. The resulting image, first published in \cite{2022Natur.602..403G}, is shown in Figure~\ref{fig:taperedContinuum}. The restoring beam is $4.3 \times 3.3$~mas, PA~$-21\fdg8$, and rms image noise is 0.16~mJy~beam\mone. The flux density of the recovered continuum is $S_{\nu} = 13.8 \pm 0.3$~mJy. The centroid position of the resolved continuum is $\mbox{RA(J2000)} = 02\mbox{h}~42\mbox{m}~40\fs70901$, $\mbox{Dec(J2000)} = -00\degr~00\arcmin~47\farcs9448$. This 22~GHz position agrees well with the VLBA position of S1 measured at 5 and 8~GHz \citep{2004ApJ...613..794G}. 

\subsection{Substructures and Filaments within the S1 Maser Spot Distribution}\label{sec:substructures}

\begin{figure*}[tbh]
  \centering
\plotone{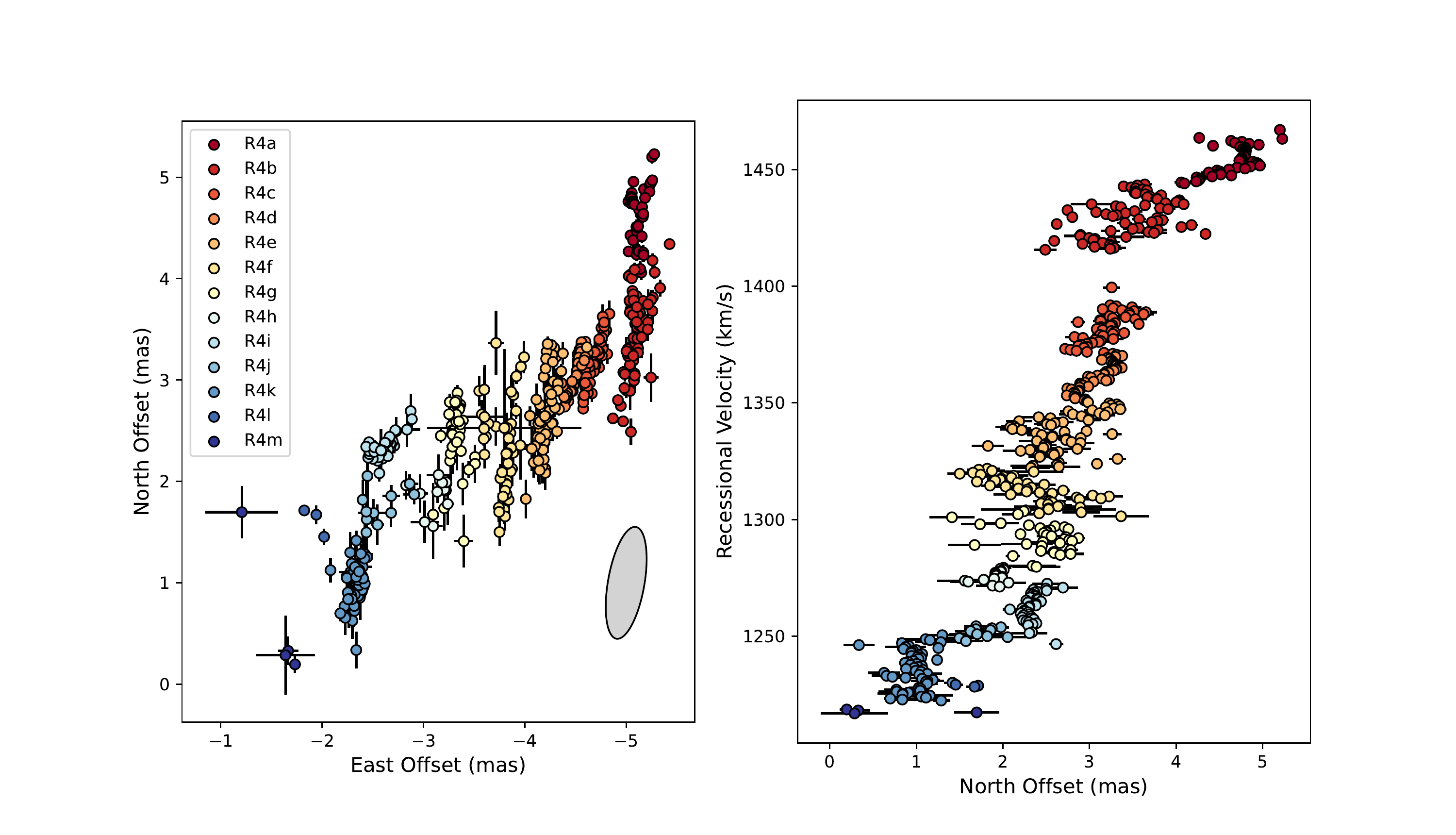}
\caption{A close-up of the R4 maser group. Subgroups are denoted by lower-case suffixes and are color-coded as shown in the legend. {\em Left panel:} the sky distribution of maser spots.  The gray-colored ellipse represents the synthetic beam. {\em Right panel:} the declination-velocity diagram of the maser spots. The subgroups form a pattern of nearly north-south, linear features that show velocity gradients ranging from about $-8$ to $+20$~\kms~mas\mone\ along the linear axis.\label{fig:spokesR4}}
\end{figure*}

GG97 reported the appearance of arcuate and linear substructures among the distribution of maser spots in the sky. As evident in Figure~\ref{fig:s1-skyplot}, we also find linear and arcuate groupings of maser spots on the sky. Shown in Fig.~\ref{fig:spokesR4}, the R4 group, in particular, breaks down into nearly parallel filaments. We identified and labeled 34 such subgroups by means of a clustering algorithm. The details of the analysis and magnified plots of the other subgroups are provided in Appendix~\ref{app:substructure}. For the purposes of discussion, we labeled each subgroup by its parent group name and a lower-case suffix increasing alphabetically to the east (e.g., R2a, R2b, R3a, R3b, etc.).

Many subgroups, particularly those belonging to group R4, orient nearly north-south, roughly 50\degr{} in PA from the overall distribution of maser spots on the sky. On the one hand, the nuclear jet and molecular outflow axes also point roughly north-south on the sky \citep{1996ApJ...464..198G, 2016ApJ...829L...7G}, so the \water{} maser filaments may be tracing gas participating in the molecular outflow. However, the synthetic beam also orients roughly north-south. The concern then is whether the apparent filaments are artifacts resulting from channel-to-channel measurement uncertainties or calibration errors. Put another way, perhaps the substructures belong to spatially unresolved clumps, but random measurement errors or calibration (or other systematic) errors introduce an apparent scatter of maser spots comparable to the size and orientation of the beam. However, such errors should not introduce significant velocity gradients within the subgroups. 

\begin{figure*}[tbh]
  \centering
\plotone{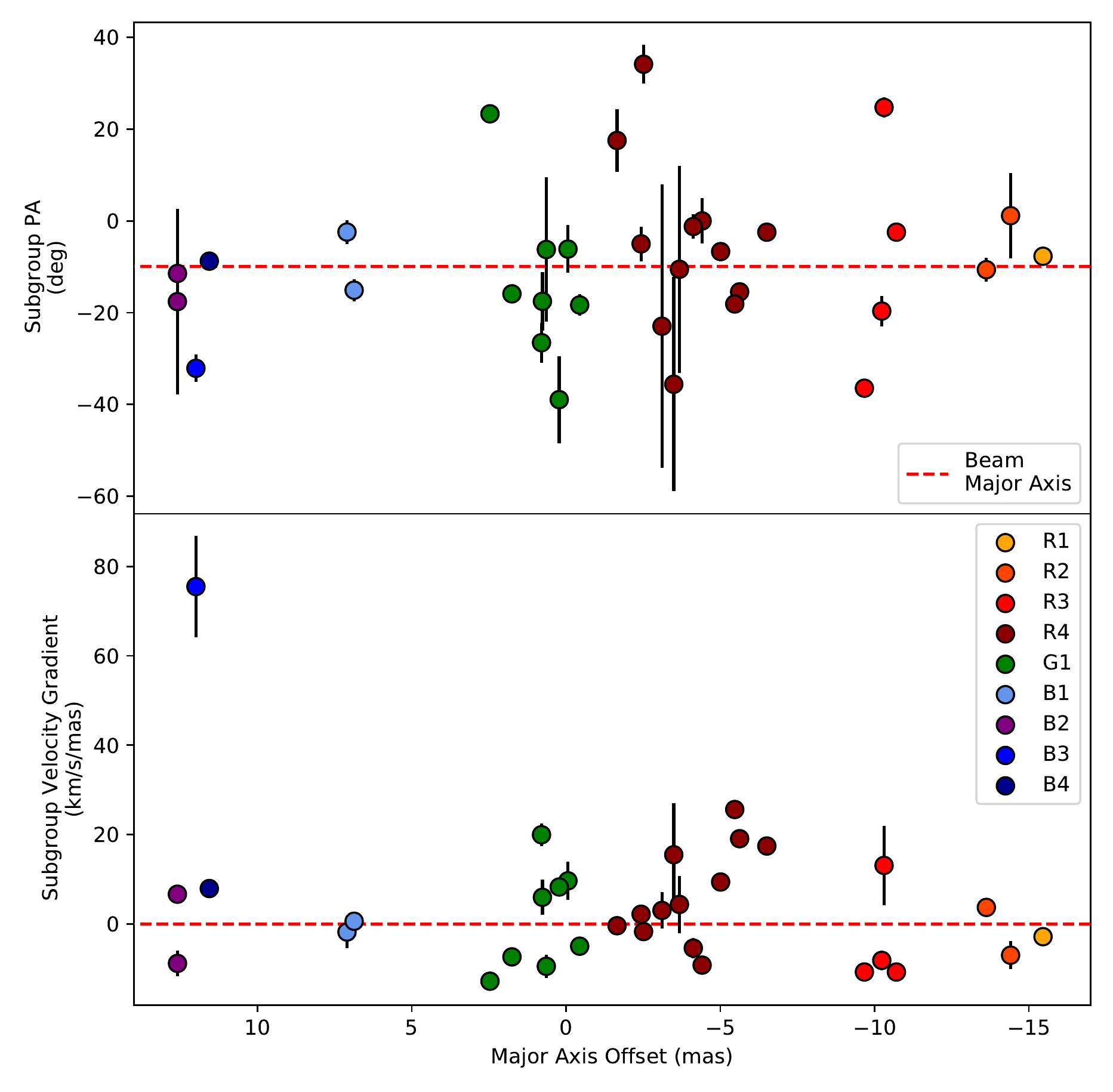}
\caption{Properties of the maser subgroups plotted vs. the major axis offset. The principal groups are color-coded as shown in the legend. The upper plot shows the position angle of each subgroup compared to the synthetic beam position angle, which is indicated by the red dashed line. The lower plot shows the mean velocity gradient along the major axis of each subgroup. The red dashed line traces zero velocity gradient. \label{fig:subgroupprops}}
\end{figure*}

In Fig.~\ref{fig:subgroupprops}, we plot the orientation and velocity gradients of the subgroups. There are nine (9) subgroups of 34 that have (1) major axis position angles within $3\sigma$ of the synthetic beam position angle and (2) velocity gradients within $3\sigma$ of zero: R2a, R4h, R4i, R4j, R4k, G1b, G1e, B1a, and B1b. Furthermore, the scatter of maser spots within these subgroups is comparable to or smaller than the size of the beam. We conclude that these subgroups, in particular, are likely unresolved. The other 25 subgroups are significantly rotated from the major axis of the beam or have significant and varying velocity gradients. Additionally, there is no consistent pattern in the velocity gradients that might otherwise hint at calibration errors. Any systematic errors should equally have affected the jet masers, for which we find no north-south (or any) filamentary structure. We conclude that most of the subgroups among the disk masers trace real substructures within the overall distribution of maser spots on the sky.

\subsection{The Position-Velocity Diagram of the Disk Masers}\label{sec:pvdiagram}
\begin{figure*}[tbh]
  \centering
\plotone{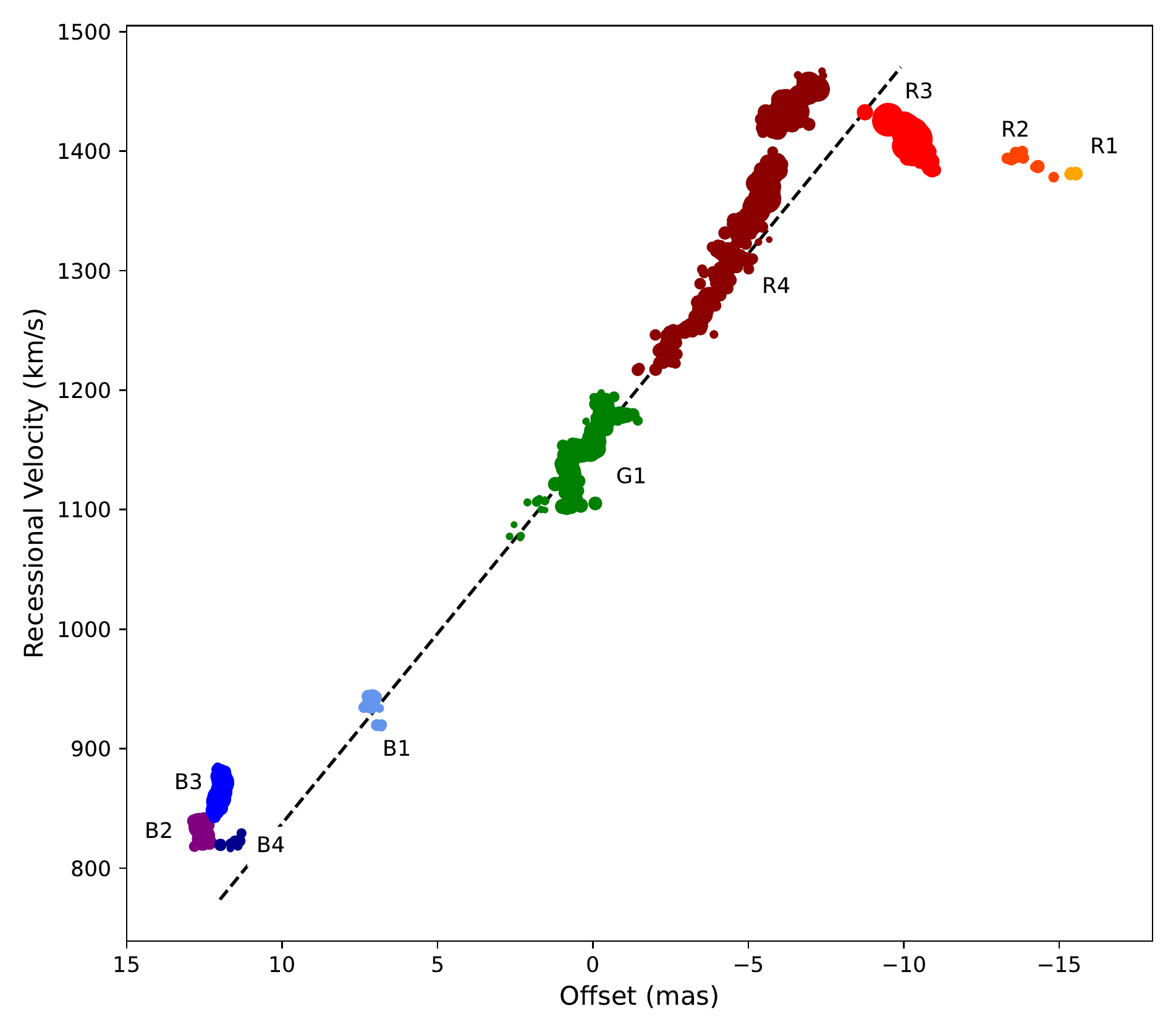}
\caption{Position-velocity diagram of the nuclear \water\ masers. Position offsets are taken along PA~$-50\degr$. The data have been colored and annotated based on the maser group label. The dashed line shows a linear fit to the G1 and B1 maser groups, extrapolated to span the observed maser velocities. The fit is not intended as a physical model but rather is intended to highlight the curvature of the distribution of maser spots in the position-velocity plane. \label{fig:pvdiagram}}
\end{figure*}

The position-velocity (\pv{}) diagram is shown in Figure~\ref{fig:pvdiagram}. To estimate the position angle of the major axis, we fit a line to the spot positions of the maser groups R4 and G1; the best-fit PA is $-50\degr\pm1\degr$. The \pv{} diagram broadly agrees with that of GG97. Groups R1--3 show falling velocities as expected for masers tracing the receding side of a rotating disk. This pattern is not matched on the blueshifted side; groups B2, B3, and B4 show a more complex arrangement in positions and velocity. 

Maser disks commonly show a linear region between the maximum velocities on the \pv{} diagram \citep[e.g.,][]{1995PNAS...9211427M, 1999Natur.400..539H,2005ARA&A..43..625L,2011ApJ...727...20K,2016ApJ...817..128G,2017ApJ...834...52G}\explain{specific references added here}. The linear region {traces} the inner {radius} of the maser region of the molecular disk {\citep[e.g.,][]{1994ApJ...432L..35W}}. Specifically, if the rotational velocity, \vrot{}, depends only on the radius of the disk $r$, then the observed recessional velocities follow $V_R = \vsys + \vrot (x / r) \sin{i}$, where \vsys{} is the systemic velocity, $x$ is the displacement on the sky along the disk midline (the offset along the projected major axis of the spot distribution), and $i$ is the inclination of the disk. For a ring of constant radius $r$, the \pv{} diagram is linear. 

The disk masers of NGC~1068, however, show a curved pattern of spots between the maximum observed velocities on the diagram \pv{}, that is, between groups R4 and B4. To illustrate, we fit a line to the \pv{} coordinates of the spots in groups G1 and B1 and extrapolated the fit to cover the range of observed maser velocities. From inspection of Figure~\ref{fig:pvdiagram}, the R4 group curves away from the best-fit line, displacing to higher recessional velocities with distance from the G1 masers. The masers of groups B2--4 also tend to higher recessional velocities than the extrapolated trend line. This apparent curvature between the R4 and B4 masers in the \pv{} diagram indicates that the orbital geometry is not a simple, rotating ring as expected for the inner radius of the \water{} maser region.

\section{Kinematic Models for the Disk Masers}\label{sec:kinematicmodels}

In this section, we consider three kinematic models to explain the apparent curvature between the velocity extremes on the \pv{} diagram (i.e., between the R4 and B4 maser groups). In the first model, the expanding ring model, radial motions (infall or outflow) contribute to the observed recessional velocities for spot groups R4--B4. The second model assumes that the R4--B4 masers follow a common elliptical orbit. In the third model, we explore the possibility that the masers arise from spiral arms in the molecular accretion disk. In all three models, we assume that the brightest maser spots originate from molecular gas on the near side of the disk midline and the continuum source. The main argument is that the continuum source provides seed radiation that is amplified by the \water{} masers. Furthermore, attenuation through ionized gas inside the molecular disk may suppress maser emission from the far side \citep[e.g.,][]{1994ApJ...432L..35W}. We describe in turn the model fitting and selection techniques (Section~\ref{sec:mcmc}), particular details of each model (Sections~\ref{sec:radialmotions}--\ref{sec:spiralArmModel}), and we provide a discussion comparing the merits of the three kinematic models (Section~\ref{sec:kinematicsDiscussion}). 

\subsection{Model Fitting Techniques}\label{sec:mcmc}

We used the Markov Chain Monte Carlo (MCMC) code PyDREAM \citep{10.1093/bioinformatics/btx626} to fit kinematic models to the maser spot data. A more complete explanation of the MCMC technique is provided in Appendix~\ref{sec:mcmcdetails}. Briefly, the algorithm generates random walks through the parameter space. For a trial set of parameters, a discretized model orbit was calculated in the disk frame and then projected onto the sky based on the model inclination ($i$), position angle ($\Omega$), and kinematic center ($X_0, Y_0$). Maser spots were then matched to the nearest point on the projected model orbit, and the posterior probability of the fit was calculated. To account for systematic uncertainty, we included the (fitted) error floors $\delta X$, $\delta Y$, and $\delta V$, which are added in quadrature to the measurement uncertainties (cf. \citealt{2013ApJ...775...13H}). At each step, a set of parameters is accepted or rejected according to the Metropolis criterion \citep{1953JChPh..21.1087M}. The integrated autocorrelation time (IAT) was used to evaluate convergence \citep{2013PASP..125..306F}.

We used a mixture model to accommodate kinematic outliers. Using the formalism of mixture models, the kinematic model is the foreground, and outliers belong to the background model. The posterior probability includes a sum of probabilities: effectively, the probability that a given maser spot belongs to the foreground and the probability that the maser spot belongs to the background. The background is modeled as a Gaussian distribution in sky coordinates and recessional velocity. Therefore, the background model introduces seven additional parameters: $P_{fg}$, the probability that a maser spot belongs to the foreground (the kinematic model under consideration); the background centroids $X_{bg}$, $Y_{bg}$, and $Z_{bg}$; and the standard deviations of the background $(\delta X_{bg},\,\delta Y_{bg},\,\delta V_{bg})$.

We used the marginal likelihood, sometimes called the {\em evidence}, for model selection. The marginal likelihood is the probability of observing the data given the model, and so models with a greater (log) marginal likelihood are favored as better representing the data. We used two estimators of the marginal likelihood, the widely applicable Bayes information criterion (WBIC) \citep{Friel2017InvestigationOT} and the thermodynamic integration estimator (TIE) \citep{10.1214/ss/1028905934,10.2307/1390653}. {The comparison of two models is summarized by the Bayes factor (BF), which is the ratio of the probability of obtaining the data under two different models: $\log{\mbox{BF}_{1,2}} =\log{[\mbox{evidence(model\,1)}] - \log{[\mbox{evidence(model\,2)}]}}$ \citep[see][for a review]{doi:10.1080/01621459.1995.10476572}.} {The estimates of the marginal likelihoods and the corresponding Bayes factors for the expanding ring, elliptical orbit, and spiral arm models} are provided in Table~\ref{tab:wbic}.

\begin{deluxetable}{lD@{$\pm$}DD@{$\pm$}DD@{$\pm$}DD@{$\pm$}D}
\tablecaption{Model Selection Statistics \explain{Included estimates of the Bayes factor relative to the best-fitting model.}\label{tab:wbic}}
\tablehead{Model & \multicolumn4c{WBIC\tablenotemark{a}} & 
\multicolumn4c{$\log{(\mbox{BF})}$\tablenotemark{b}} & \multicolumn4c{TIE\tablenotemark{c}} & 
\multicolumn4c{$\log{(\mbox{BF})}$\tablenotemark{b}}}
\decimals
\startdata
Expanding Ring   & $-4887.3$   & 0.4 & $-1283$ & 1 & $-4635$ & 1 & $-1109$ & 1 \\ 
Elliptical Orbit & $-4484$   & 3 & $-880$ & 3 & $-4473$ & 2 & $-947$ & 2 \\ 
Spiral Arms       & $-3604$   & 1 & \multicolumn{4}{c}{\nodata} & $-3526$ & 1 & \multicolumn{4}{c}{\nodata} \\ 
\enddata
\tablenotetext{a}{The widely applicable Bayes information criterion. Larger (more positive) values indicate that the model is more likely to produce the observed data.}
\tablenotetext{b}{An estimate of the Bayes factor (BF): here, the ratio of the probability of obtaining the data for the given model to the same probability for the spiral arms model.}
\tablenotetext{c}{The marginal likelihood estimated by thermodynamic integration. Again, larger (more positive) values indicate the preferred model.}
\end{deluxetable}

\subsection{Expanding Ring Model}\label{sec:radialmotions}

\begin{figure*}[tbh]
  \centering
\includegraphics[width=\textwidth]{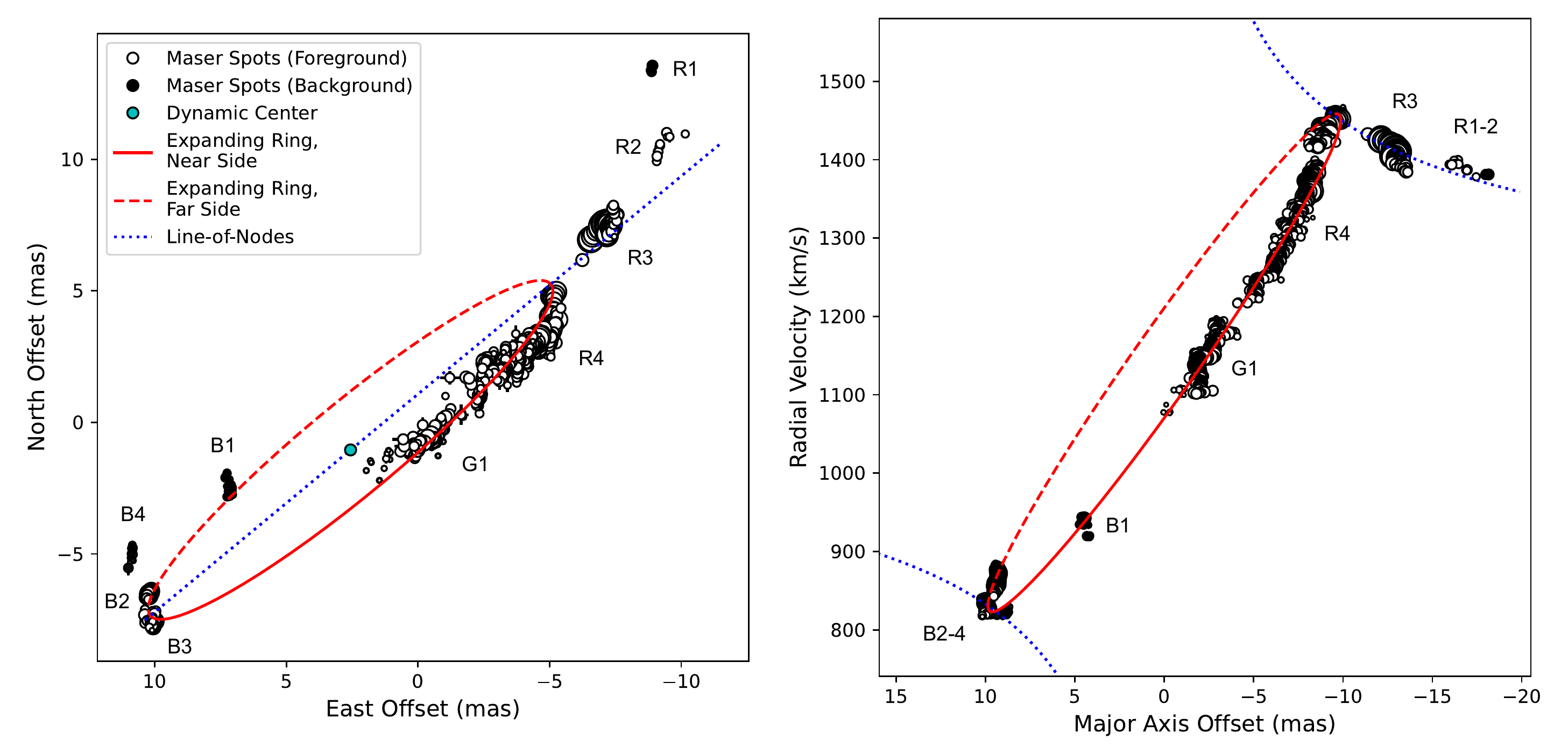}
\caption{A representative fit of the expanding ring model, in sky coordinates, for the maser disk of NGC~1068. In both panels, the maser spot positions and velocities are shown as shaded circles. The circle diameters are scaled to the flux of the maser spots. The shading depends on the foreground probability for individual spots: 
lighter shades indicate maser spots that are likely part of the model orbit (the foreground model), and darker shades indicate outliers.
Left panel: sky plot of best-fit circular orbit, traced by the red lines; the near side of the midline is plotted as a solid line, and the far side as a dashed line. The kinematic center is shown as a cyan circle. The blue dotted line traces the disk midline. Right panel: the position-velocity diagram. The blue dotted lines trace the Keplerian circular speed curve.\label{fig:ringFit}}
\end{figure*}

In this model, we assume that the R4--B4 maser spots occupy a narrow annulus of radius $r_0$ and the R1--R3 maser spots fall along the disk midline.  The kinematics are defined by constant radial velocity $v_r$ (representing infall or outflow) and constant azimuthal (i.e. rotational) velocity $v_0$. To summarize, for a single circular orbit with uniform radial motion, there are eleven (11) parameters in the foreground model: $X_0$, $Y_0$, \vsys{}, $r_0$, $v_r$, $v_0$, $i$, $\Omega$, and error floors $\delta X$, $\delta Y$, and $\delta V$. The results of the expanding ring model are provided in Table~\ref{tab:outflowOrbit} and illustrated in Figure~\ref{fig:ringFit}. 

\begin{deluxetable}{llD@{$\pm$}Dlr}
\tablecaption{Expanding Ring Model\label{tab:outflowOrbit}}
\tablehead{Parameter & Prior\tablenotemark{a} & \multicolumn4c{Value} &  Units & $N/{\rm IAT}$}
\decimals
\startdata
\multicolumn{8}{c}{Foreground Model}\\
\hline
Foreground Prob., $P_{fg}$ & $\mathcal{U}(0,1)$ & 0.954 & 0.007 & \nodata & 5600\\
Kinematic center, $X_0$ & $\mathcal{U}(-10, 10)$ & 2.56 & 0.01 & mas & 5180\\
Kinematic center, $Y_0$ & $\mathcal{U}(-10, 10)$ & $-1.05$ & 0.03 & mas & 4829\\
Systemic velocity, \vsys{} & $\mathcal{N}(1132, 5)$& 1140.8 & 0.5 & \kms{} & 5117\\
Ring radius, $r_0$   & $\mathcal{U}(5, 25)$& 9.87 & 0.02 & mas & 4898\\
Ring rotation speed, $v_0$ & $\mathcal{U}(300, 500)$ & 314.9 &  0.5 & \kms{} & 5626 \\
Outflow speed, $v_r$ & $\mathcal{U}(-100, 100)$& 72 & 1 & \kms{} & 4464  \\
Inclination, $i$ & $\mathcal{U}(90, 110)$& 79.9 & 0.2 & degrees & 4584 \\
Position Angle, $\Omega$ & $\mathcal{U}(270, 360)$& 309.6 & 0.1 & degrees & 5125 \\
Systematic error, $\delta X$ & $\mathcal{U}(0, 4)$ & 0.217 & 0.006 & mas & 5383 \\
Systematic error, $\delta Y$ & $\mathcal{U}(0, 4)$ & 0.57 & 0.02 & mas & 5681 \\
Systematic error, $\delta V$ & $\mathcal{U}(0, 20)$& 4.3 & 0.1 & \kms{} & 5954 \\
\hline
\multicolumn{8}{c}{Background Model}\\
\hline
Center, $X_{bg}$ & $\mathcal{U}(-20,20)$ & 6.1 & 0.8 & mas & 6039 \\
Center, $Y_{bg}$ & $\mathcal{U}(-20,20)$ & $-1.3$ & 0.8 & mas & 5281 \\
Center, $V_{bg}$ & $\mathcal{N}(1132,400)$ & 960 & 20 & \kms{} & 5916\\
Width, $\delta X_{bg}$ & $\mathcal{H}(0,20)$ & 6.1 & 0.8 & mas & 6039\\
Width, $\delta Y_{bg}$ & $\mathcal{H}(0,20)$ & $-1.3$ & 0.8 & mas & 5883 \\
Width, $\delta V_{bg}$ & $\mathcal{H}(0,100)$& 150 & 20 & \kms{} & 5426 \\
\hline
\multicolumn{8}{c}{Derived Parameters\tablenotemark{b}}\\
\hline
Ring radius & \nodata & 0.670 & 0.001 & pc & \nodata \\
Central Mass & \nodata & 15.45 & 0.06 & $10^6$ \Mo{} & \nodata \\
Orbital Period at $r=r_0$ & \nodata & 13.07 & 0.03 & kyr & \nodata \\
\enddata
\tablenotetext{a}{$\mathcal{U}(x_l, x_h)$ is a uniform distribution with lower bound $x_l$ and upper bound $x_h$. $\mathcal{N}(\mu, \sigma)$ is a normal distribution with mean $\mu$ and standard deviation $\sigma$. $\mathcal{H}(\mu, \sigma)$ is a half-normal distribution with lower limit $\mu$.}
\tablenotetext{b}{Assumes $1\arcsec = 70$~pc.}
\end{deluxetable}

\subsection{Elliptical Orbit Model}\label{sec:ellipticalOrbit}

\begin{figure*}[tbh]
  \centering
\includegraphics[width=\textwidth]{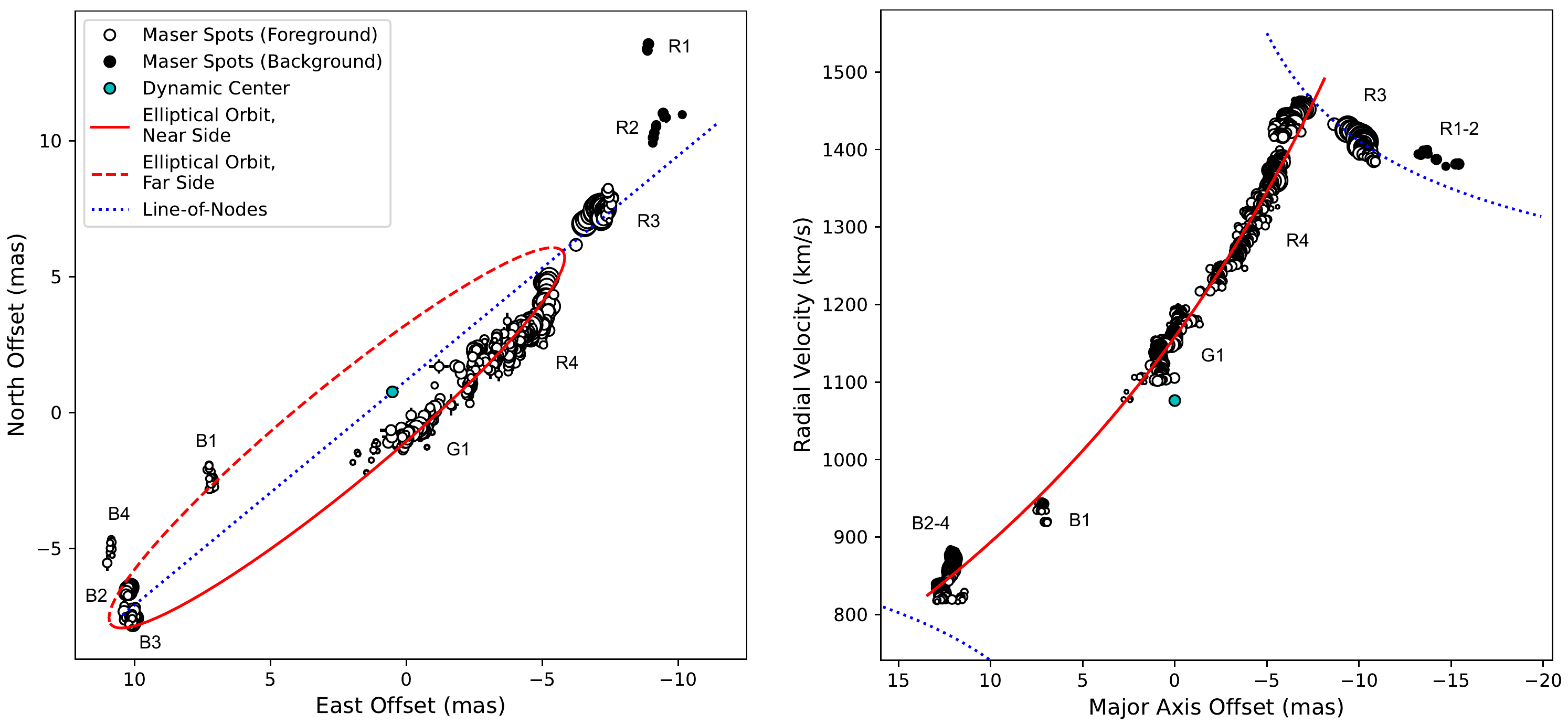}
\caption{A representative fit of the elliptical orbit model, in sky coordinates, for the maser disk of NGC~1068. The plotting conventions are as in Fig.~\ref{fig:ringFit}. \label{fig:ellipticalFit}}
\end{figure*}

As in the expanding-ring model, the elliptical orbit model is a single orbit model applied to maser groups R4 -- B4. In addition to the sky projection angles $i$ and $\Omega$, 
elliptical orbits require an additional angle, $\omega$, the argument of periapsis, which is the azimuthal angle between midline and the periapsis (cf. \citealt{2013ApJ...775...13H}). The shape of the orbit is determined by the semimajor axis $a$, and the eccentricity \ecc. In cylindrical coordinates ($r$, $\phi$), the shape of the orbit is given by
\begin{equation}
r = \frac{a (1 - \ecc^2)}{1 + \ecc \cos(\phi - \omega)}\,.
\end{equation}
In the disk rest frame, the radial and azimuthal components of the orbital velocity, $v_r$ and $v_\phi$, are
\begin{eqnarray*}
v_r & = & v_a\,\frac{\ecc \sin(\phi - \omega)}{\sqrt{1 - \ecc^2}}, \,\mbox{and}  \\
v_{\phi} & = & v_a\,\frac{1 + \ecc\cos(\phi - \omega)}{\sqrt{1 - \ecc^2}}\ , 
\end{eqnarray*}
where $v_a$ is the Keplerian circular speed at radius $a$. The foreground model requires twelve (12) parameters: $X_0$, $Y_0$, \vsys{}, $a$, $\ecc$, $v_a$, $\omega$, $i$, $\Omega$, $\delta X$, $\delta Y$, $\delta V$. The results of the fit are provided in Table~\ref{tab:ellipticalOrbit} and Fig.~\ref{fig:ellipticalFit}. 

\begin{deluxetable}{llD@{$\pm$}Dlr}
\tablecaption{Elliptical Orbit Model\label{tab:ellipticalOrbit}}
\tablehead{Parameter & Prior\tablenotemark{a} & \multicolumn4c{Value} &  Units & $N/{\rm IAT}$}
\decimals
\startdata
\multicolumn{8}{c}{Foreground Model}\\
\hline
Foreground Prob., $P_{fg}$ & $\mathcal{U}(0,1)$ & 0.984 & 0.006 & \nodata & 4865\\
Kinematic center, $X_0$ & $\mathcal{U}(-10, 10)$ & 0.49 & 0.04 & mas & 3973\\
Kinematic center, $Y_0$ & $\mathcal{U}(-10, 10)$ & 0.74 & 0.03 & mas & 4021\\
Systemic velocity, \vsys{} & $\mathcal{N}(1132, 5)$& 1070 & 1 & \kms{} & 4043\\
Semimajor axis, $a$   & $\mathcal{U}(5, 25)$& 10.95 & 0.05 & mas & 3919\\
Circular speed at $a$, $v_{a}$ & $\mathcal{U}(300, 500)$ & 330.3 &  0.9 & \kms{} & 4060 \\
Eccentricity, $\ecc$ & $\mathcal{U}(0, 1)$& 0.258 & 0.004 & \ldots & 4235  \\
Argument of periapsis, $\omega$ & $\mathcal{U}(0,360)$& 0.07 & 0.08 & degrees & 2599  \\
Inclination, $i$ & $\mathcal{U}(90, 110)$& 80.7 & 0.1 & degrees & 4193 \\
Position Angle, $\Omega$ & $\mathcal{U}(270, 360)$& 309.5 & 0.1 & degrees & 3967 \\
Systematic error, $\delta X$ & $\mathcal{U}(0, 4)$ & 0.228 & 0.005 & mas & 4204 \\
Systematic error, $\delta Y$ & $\mathcal{U}(0, 4)$ & 0.58 & 0.01 & mas & 3915 \\
Systematic error, $\delta V$ & $\mathcal{U}(0, 20)$& 4.7 & 0.1 & \kms{} & 4449 \\
\hline
\multicolumn{8}{c}{Background Model}\\
\hline
Center, $X_{bg}$ & $\mathcal{U}(-20,20)$ & $-9.19$ & 0.09 & mas & 4129 \\
Center, $Y_{bg}$ & $\mathcal{U}(-20,20)$ & $11.5$ & 0.4 & mas & 4171 \\
Center, $V_{bg}$ & $\mathcal{N}(1132,400)$ & 1388 & 2 & \kms{} & 4487\\
Width, $\delta X_{bg}$ & $\mathcal{H}(0,20)$ & 0.36 & 0.07 & mas & 3733\\
Width, $\delta Y_{bg}$ & $\mathcal{H}(0,20)$ & 1.5 & 0.3 & mas & 3221 \\
Width, $\delta V_{bg}$ & $\mathcal{H}(0,100)$& 8 & 2 & \kms{} & 3694 \\
\hline
\multicolumn{8}{c}{Derived Parameters\tablenotemark{b}}\\
\hline
Semimajor axis & \nodata & 0.744 & 0.004 & pc & \nodata \\
Central Mass & \nodata & 18.9 & 0.2 & $10^6$ \Mo{} & \nodata \\
Orbital Period & \nodata & 13.83 & 0.04 & kyr & \nodata \\
\enddata
\tablenotetext{a}{$\mathcal{U}(x_l, x_h)$ is a uniform distribution with lower bound $x_l$ and upper bound $x_h$. $\mathcal{N}(\mu, \sigma)$ is a normal distribution with mean $\mu$ and standard deviation $\sigma$. $\mathcal{H}(\mu, \sigma)$ is a half-normal distribution with lower limit $\mu$.}
\tablenotetext{b}{Assumes $1\arcsec = 70$~pc.}
\end{deluxetable}

\subsection{Spiral Arms}\label{sec:spiralArmModel}

In the spiral arms model, the \water{} masers trace molecular gas at a smoothly changing distance from the dynamical center, which introduces the curve on the \pv{} diagram between the maser groups R4 and B4. We adopted a logarithmic spiral model for the distribution of maser spots, 
\begin{equation}
r = r_0 \exp\left(\phi \tan \theta_P \right),
\end{equation}
where $r_0$ is a length scale parameter and $\theta_P$ is the pitch angle of the arms. We further assume that there are two symmetric arms, a main arm intended to fit the R4--G1 masers and a symmetric opposite arm rotated by $\Delta\phi=180\degr{}$ in the disk plane. 

To simplify the model, we assume that Keplerian rotation dominates the kinematics, but we also allow for uniform streaming motion locally parallel to the arms. The inclusion of streaming along the arm introduces an additional parameter, the streaming speed $v_S$. Radial and azimuthal velocities in the disk frame are given by 
\begin{equation}
\begin{split}
v_r      & =  v_S\,\sin{\theta_P},\, \mbox{and} \\
v_{\phi} & =  v_0\,\sqrt{\frac{r_0}{r}} + v_S\,\cos{\theta_P},
\end{split}\label{eqn:spiralmotion}
\end{equation}
where $v_0$ is the Keplerian rotation speed at $r=r_0$. 

\begin{figure*}[tbh]
  \centering
\includegraphics[width=\textwidth]{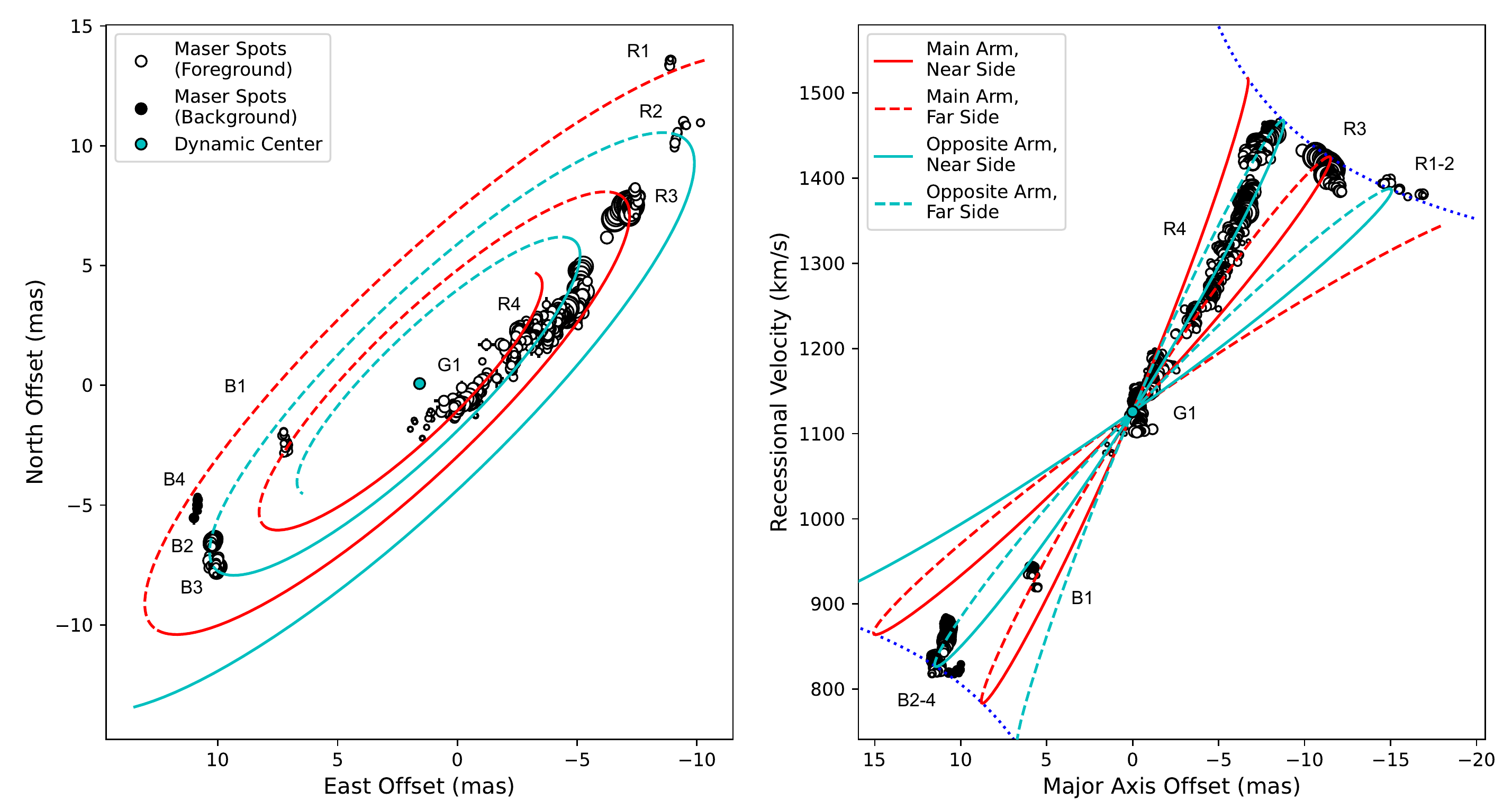}
\caption{A representative fit of the spiral arms model, in sky coordinates, for the maser disk of NGC~1068. In both panels, the main arm is traced by a red line, and the opposite arm is traced by a cyan line. Otherwise, the plotting conventions are as in Fig.~\ref{fig:ringFit}. \label{fig:spiralFit}}
\end{figure*}

In summary, the spiral arm model requires twelve (12) parameters: $x_0$, $y_0$, $r_0$, \vsys, $v_0$, $v_S$, $\theta_P$, $i$, $\Omega$, $\delta X$, $\delta Y$, $\delta V$. The results are summarized in Tables~\ref{tab:spiralArm} and plotted in Figure~\ref{fig:spiralFit}.

\begin{deluxetable}{llD@{$\pm$}Dlr}
\tablecaption{Spiral Arms Model\label{tab:spiralArm}}
\tablehead{Parameter & Prior\tablenotemark{a} & \multicolumn4c{Value} &  Units & $N/{\rm IAT}$}
\decimals
\startdata
\multicolumn{8}{c}{Foreground Model}\\
\hline
Foreground Prob., $P_{fg}$ & $\mathcal{U}(0,1)$ & 0.990 & 0.003 & \nodata & 2699\\
Kinematic center, $X_0$ & $\mathcal{U}(-10, 10)$ & 1.579 & 0.007 & mas & 2472\\
Kinematic center, $Y_0$ & $\mathcal{U}(-10, 10)$ & 0.07 & 0.02 & mas & 2180\\
Systemic velocity, \vsys{} & $\mathcal{N}(1132, 5)$& 1125.8 & 0.2 & \kms{} & 2174\\
Scale radius, $r_0$   & $\mathcal{U}(5, 25)$& 6.70 & 0.02 & mas & 2711\\
Circular speed at $r_0$, $v_0$ & $\mathcal{U}(300, 500)$ & 404 &  1 & \kms{} & 1688 \\
Streaming speed, $v_S$  & $\mathcal{U}(0, 100)$ & 0.4 &  0.5 & \kms{} & 1439 \\
Pitch angle, $\theta_P$ & $\mathcal{U}(4,24)$& 4.89 & 0.03 & degrees & 2426  \\
Inclination, $i$ & $\mathcal{U}(90, 110)$& 75.5 & 0.1 & degrees & 1856 \\
Position Angle, $\Omega$ & $\mathcal{U}(270, 360)$& 313.4 & 0.1 & degrees & 1975 \\
Systematic error, $\delta X$ & $\mathcal{U}(0, 4)$ & 0.153 & 0.004 & mas & 2508 \\
Systematic error, $\delta Y$ & $\mathcal{U}(0, 4)$ & 0.74 & 0.02 & mas & 2873 \\
Systematic error, $\delta V$ & $\mathcal{U}(0, 20)$& 2.60 & 0.06 & \kms{} & 2699 \\
\hline
\multicolumn{8}{c}{Background Model}\\
\hline
Center, $X_{bg}$ & $\mathcal{U}(-20,20)$ & 10.87 & 0.02 & mas & 2014 \\
Center, $Y_{bg}$ & $\mathcal{U}(-20,20)$ & $-4.99$ & 0.07 & mas & 2404 \\
Center, $V_{bg}$ & $\mathcal{N}(1132,400)$ & 821 & 1 & \kms{} & 2276 \\
Width, $\delta X_{bg}$ & $\mathcal{H}(0,20)$ & 0.04 & 0.02 & mas & 1823\\
Width, $\delta Y_{bg}$ & $\mathcal{H}(0,20)$ & 0.19 & 0.07 & mas & 2218 \\
Width, $\delta V_{bg}$ & $\mathcal{H}(0,100)$& 4 & 1 & \kms{} & 1779 \\
\hline
\multicolumn{8}{c}{Derived Parameters\tablenotemark{b}}\\
\hline
Scale radius & \nodata & 0.455 & 0.002 & pc & \nodata \\
Central Mass & \nodata & 17.2 & 0.1 & $10^6$ \Mo{} & \nodata \\
Orbital Period at $r=r_0$ & \nodata & 6.92 & 0.04 & kyr & \nodata \\
\enddata
\tablenotetext{a}{$\mathcal{U}(x_l, x_h)$ is a uniform distribution with lower bound $x_l$ and upper bound $x_h$. $\mathcal{N}(\mu, \sigma)$ is a normal distribution with mean $\mu$ and standard deviation $\sigma$. $\mathcal{H}(\mu, \sigma)$ is a half-normal distribution with lower limit $\mu$.}
\tablenotetext{b}{Assumes $1\arcsec = 70$~pc.}
\end{deluxetable}

\subsection{Kinematic Model Selection\label{sec:kinematicsDiscussion}}

\explain{We renamed this subsection in response to the addition of a new subsection, below.}

Overall, the spiral arms model better fits the data compared to the expanding ring and elliptical orbit model. Formally, the spiral arms model has the highest marginal likelihood by a factor of $\sim \exp(1000)$ relative to the other models (Table~\ref{tab:wbic}). However, this result should be viewed critically because, from inspection of Figs.~\ref{fig:ringFit} and \ref{fig:ellipticalFit}, the disk midline model for groups R1--R3 is partly responsible for the poorer fits. To that point, the R1 masers are identified as outliers in both the expanding-ring and elliptical orbit models, and the R2 masers are also identified as outliers in the elliptical orbit model. Even so, both the expanding ring and elliptical orbit models are problematic in other ways. First, the expanding-ring model also fails to fit the B1 and B4 masers, which are identified as outliers. Second, the best-fit systemic velocity of the elliptical orbit model is $1070\pm1$~\kms{}, roughly 60~\kms{} blueshifted relative to the systemic velocity of the host galaxy. For comparison, the spiral arm model produces a best-fit systemic velocity within $2\sigma$ of the recessional velocity of the host galaxy and the surrounding molecular disk. Only the B4 maser group is identified as an outlier in the spiral arms model, although it appears that the R1 masers are also poorly fitted. Insofar as the spiral arms model provides, formally, the best goodness-of-fit, the best match to the systemic velocity of the host galaxy, and the best fit to most of the maser spots, we maintain the conclusion that the spiral arms model provides a better description of the data.

In contrast to previous interpretations of the \pv{} diagram, we have assumed a Keplerian rotation curve in our kinematic models. To assess this assumption, we modified the spiral arms model to include a power law rotation curve: $v_{\phi} = v_0 (r / r_0)^{\alpha}$, where $\alpha$ is a fitted parameter. As earlier interpretations concluded $\alpha \sim -0.3$ \citep{LB03}, we adopted a generous and uniform prior, $\alpha \sim \mathcal{U}(-1,0)$. The results are provided in Table~\ref{tab:spiralArm2}. For this modified spiral arm model, $\mbox{WBIC} = -3607\pm 2$ and $\mbox{TIE}=-3538\pm 2$, indicating a fit goodness comparable to the unmodified model. The best-fit power-law index is $\alpha = -0.51 \pm 0.01$, consistent with Keplerian rotation. Based on a comparison of Tables~\ref{tab:spiralArm} and \ref{tab:spiralArm2}, the other parameters do not show significant changes. 

\begin{deluxetable}{llD@{$\pm$}Dlr}
\tablecaption{Modified Spiral Arms Model\label{tab:spiralArm2}}
\tablehead{Parameter & Prior\tablenotemark{a} & \multicolumn4c{Value} &  Units & $N/{\rm IAT}$}
\decimals
\startdata
\multicolumn{8}{c}{Foreground Model}\\
\hline
Foreground Prob., $P_{fg}$ & $\mathcal{U}(0,1)$ & 0.990 & 0.003 & \nodata & 2744\\
Kinematic center, $X_0$ & $\mathcal{U}(-10, 10)$ & 1.581 & 0.007 & mas & 2165\\
Kinematic center, $Y_0$ & $\mathcal{U}(-10, 10)$ & 0.0 8 & 0.03 & mas & 2172\\
Systemic velocity, \vsys{} & $\mathcal{N}(1132, 5)$& 1125.8 & 0.3 & \kms{} & 1912\\
Scale radius, $r_0$   & $\mathcal{U}(5, 25)$& 6.71 & 0.03 & mas & 2325\\
Circular speed at $r_0$, $v_0$ & $\mathcal{U}(300, 500)$ & 406 &  2 & \kms{}  & 2345 \\
Rotation curve power-law index, $\alpha$ & $\mathcal{U}(-1,0)$ & $-0.51$ & $0.01$ & \nodata & 2468 \\
Streaming speed, $v_S$  & $\mathcal{U}(0, 100)$ & 0.5 &  0.5 & \kms{} & 1649 \\
Pitch angle, $\theta_P$ & $\mathcal{U}(4,24)$& 4.88 & 0.03 & degrees & 2289  \\
Inclination, $i$ & $\mathcal{U}(90, 110)$& 75.5 & 0.1 & degrees & 1839 \\
Position Angle, $\Omega$ & $\mathcal{U}(270, 360)$& 313.4 & 0.1 & degrees & 1839 \\
Systematic error, $\delta X$ & $\mathcal{U}(0, 4)$ & 0.153 & 0.004 & mas & 2420 \\
Systematic error, $\delta Y$ & $\mathcal{U}(0, 4)$ & 0.73 & 0.02 & mas & 2471 \\
Systematic error, $\delta V$ & $\mathcal{U}(0, 20)$& 2.59 & 0.06 & \kms{} & 2859 \\
\hline
\multicolumn{8}{c}{Background Model}\\
\hline
Center, $X_{bg}$ & $\mathcal{U}(-20,20)$ & 10.87 & 0.02 & mas & 2224 \\
Center, $Y_{bg}$ & $\mathcal{U}(-20,20)$ & $-5.00$ & 0.07 & mas & 2188 \\
Center, $V_{bg}$ & $\mathcal{N}(1132,400)$ & 821 & 1 & \kms{} & 2513 \\
Width, $\delta X_{bg}$ & $\mathcal{H}(0,20)$ & 0.04 & 0.02 & mas & 1738\\
Width, $\delta Y_{bg}$ & $\mathcal{H}(0,20)$ & 0.19 & 0.07 & mas & 2018 \\
Width, $\delta V_{bg}$ & $\mathcal{H}(0,100)$& 4 & 1 & \kms{} & 1726 \\
\hline
\multicolumn{8}{c}{Derived Parameters\tablenotemark{b}}\\
\hline
Scale radius & \nodata & 0.456 & 0.002 & pc & \nodata \\
Central Mass & \nodata & 17.4 & 0.2 & $10^6$ \Mo{} & \nodata \\
Orbital Period at $r=r_0$ & \nodata & 6.90 & 0.04 & kyr & \nodata \\
\enddata
\tablenotetext{a}{$\mathcal{U}(x_l, x_h)$ is a uniform distribution with lower bound $x_l$ and upper bound $x_h$. $\mathcal{N}(\mu, \sigma)$ is a normal distribution with mean $\mu$ and standard deviation $\sigma$. $\mathcal{H}(\mu, \sigma)$ is a half-normal distribution with lower limit $\mu$.}
\tablenotetext{b}{Assumes $1\arcsec = 70$~pc.}
\end{deluxetable}

\begin{figure*}[tbh]
  \centering
\includegraphics[width=\textwidth]{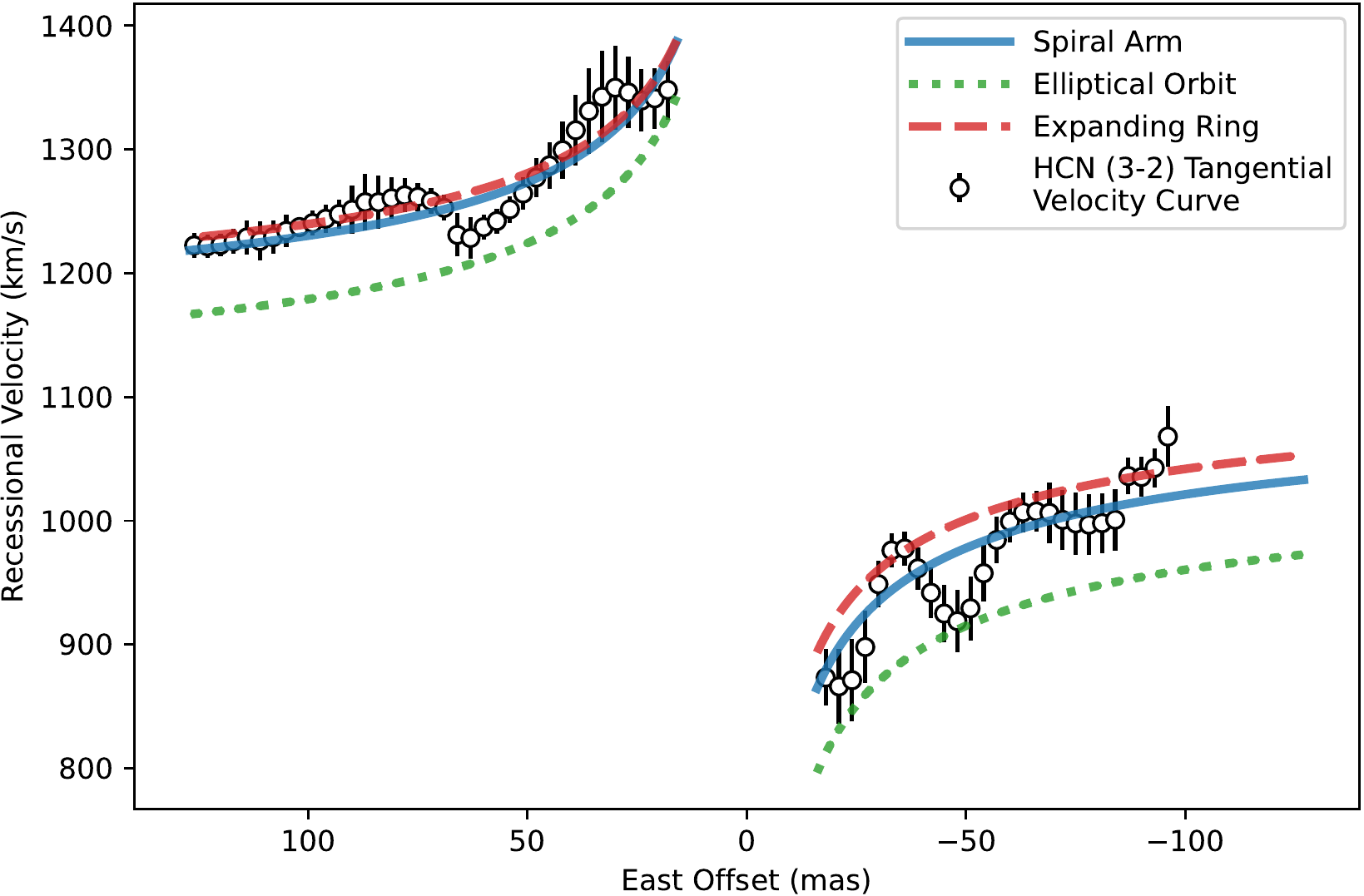}
\caption{A comparison of the extrapolated rotation curves of the \water{} maser kinematic models \water{} and the tangential velocity curve of HCN ($J=3\rightarrow 2$) tangential velocity curve from \cite{2019ApJ...884L..28I}. Note that the gas traced by HCN counterrotates relative to the \water{} maser disk, and so the rotation curves have been inverted. 
\label{fig:hcnrotcur}}
\end{figure*}

To assess the fit of the \water{} maser rotation curve further, we compare the extrapolated model rotation curves with the tangential velocity curve of HCN ($J=3\rightarrow2$) emission measured by \cite{2019ApJ...884L..28I}. The results are shown in Fig.~\ref{fig:hcnrotcur}. Note that the molecular gas traced by the HCN emission counterrotates relative to the \water{} maser disk, so the extrapolated rotation curves have been inverted. The HCN tangential velocity curve traces the kinematics on scales larger than the \water{} maser disk, $\sim 20$--100~mas ($\sim 1.4$--7~pc). Therefore, if the mass of the maser disk is a substantial fraction of the enclosed mass (see, e.g., \citealt{LB03}), the HCN velocities should exceed the extrapolated rotation curves. On the contrary, the extrapolated rotation curve of the spiral arm model agrees well with the HCN tangential velocity curve. This result leads to several important conclusions. Most obviously, the spiral arm model better predicts the HCN tangential velocity curve, further supporting it as a better model compared to the expanding ring and elliptical orbit models. Second, the close agreement between the HCN tangential velocity curve and the extrapolated \water{} maser rotation curve is likely not coincidental. Rather, it appears that counterrotation dominates the observed tangential velocities of the outer molecular disk. Third, the rotation curve of the \water{} maser disk of NGC~1068 is almost certainly Keplerian. A flatter rotation curve would extrapolate to rotation velocities that exceed those observed in the outer molecular disk. Putting these conclusions together, within a projected distance of $\sim 100$~mas (7~pc) of the kinematic center, the gravitational potential is dominated by a compact mass of $(17.2\pm 0.1)\times10^6$~\Mo{}. 

\subsection{An Examination of the Keplerian Rotation Curve}
\explain{We introduced a new subsection at the recommendation of the referee.}

An important question follows: why do the kinematic models favor Keplerian rotation while the \pv{} diagram suggests sub-Keplerian rotation? It should be emphasized that the \pv{} diagram analysis uses projections of the redshifted maser spot coordinates, only about 12\% of the data, to estimate the rotation curve, but our kinematic models use all the maser spots. Unlike spots confined to a single radius, maser spots within spiral arms sample a range of radii and help constrain the rotation curve. More specifically, the redshifted masers sample $r$ between approximately 9 and 15 mas (0.6 to 1~pc), and the spiral arms sampled by groups R4, B2 and B3 redundantly sample the range 9 to 12~mas (0.6 to 0.8~pc). Streaming motions associated with spiral arms might affect the \pv{} diagram analysis; however, we find that the streaming motions are negligible compared to the rotation speeds, $v_S = 0.4\pm 0.5$~\kms{} (Table~\ref{tab:spiralArm}). Rather, the answer lies in the assumption made in the interpretation of the \pv{} diagram, namely, that the maser groups R1--3 and the highest velocity masers of R4 lie along the disk midline. Certainly, that is a reasonable assumption as enhanced maser self-amplification is expected along the midline of an edge-on disk \citep{1994ApJ...432L..35W}. However, the plane of the maser disk is not viewed edge-on, but is tilted by nearly 15\degr{} from the line-of-sight ($i = 105\degr{}$;  Table~\ref{tab:spiralArm}), which reduces the path length of the line of sight through the molecular disk. It seems more likely then that the masers also sample local density enhancements within the disk rather as well as favorable coherent path lengths through the disk.

\begin{figure*}[tbh]
  \centering
\includegraphics[width=\textwidth]{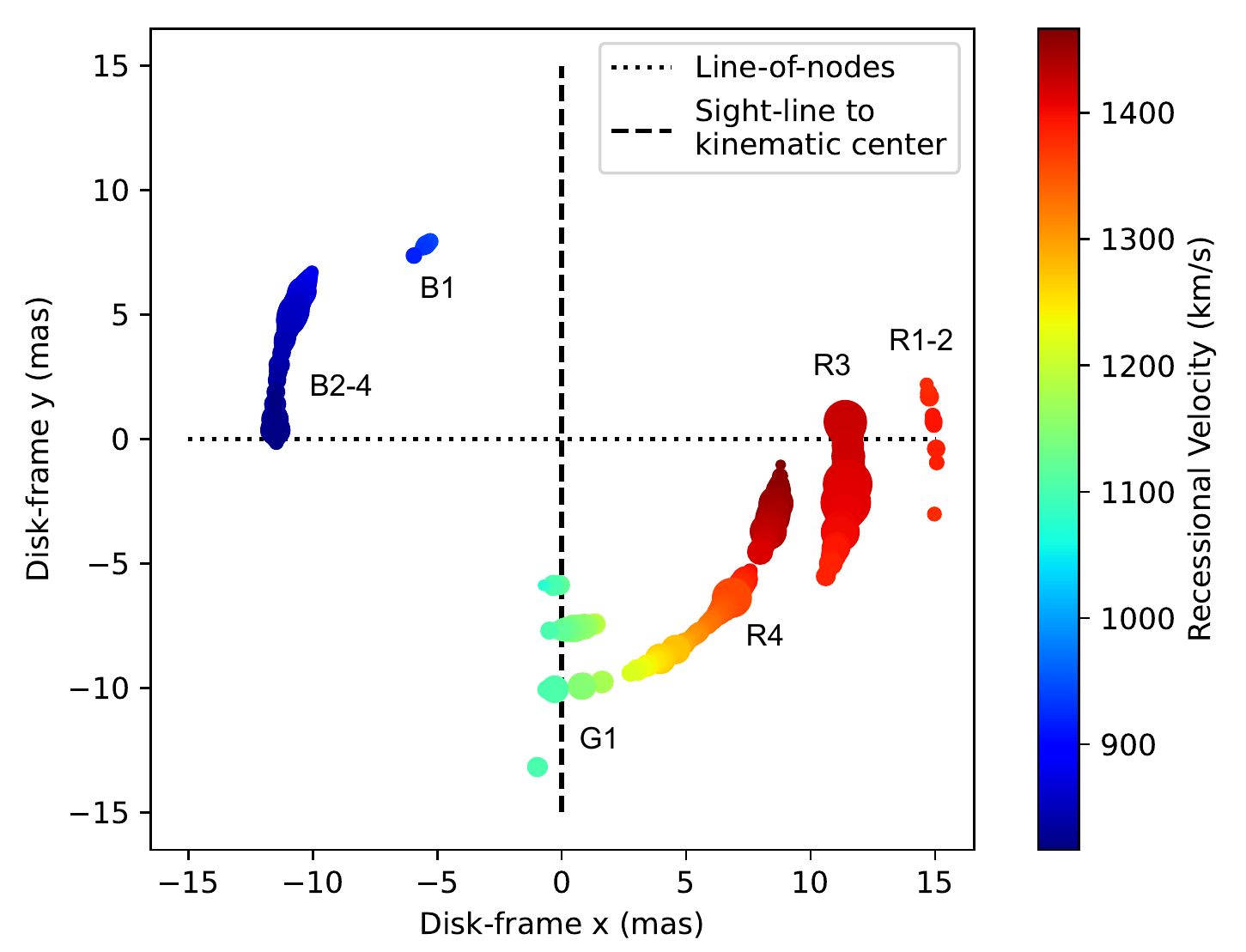}
\caption{The spiral arm model in the frame of the maser disk. The line of sight is towards positive $y$. Note that the G1 masers appear to sample arms at a range of distances, but this result is likely an artifact of projection of the near-side maser spots onto the flat disk model. \label{fig:spiralArmDeprojected}}
\end{figure*}

Fig.~\ref{fig:spiralArmDeprojected} shows the spiral arm model in the disk frame. We caution that the G1 masers suffer deprojection artifacts because, along the sight line to the kinematic center, the recessional velocity is insensitive to the rotation curve, and the deprojection of the nearly edge-on disk produces degenerate solutions in disk radius. Putting aside the deprojection of the G1 maser spots, two results stand out. First, the model favors displacing the highest-velocity masers (within the R4 masers) closer to the observer than the disk midline, the R3 masers just crossing the midline, and the R2 masers roughly symmetrically distributed across the midline. As a result, the recessional velocities sampled by the R3 and R4 masers tend to fall just below the circular speed rotation curve. The displacement especially of the high-velocity R4 masers away from the midline creates the illusion of a flatter rotation curve on the \pv{} diagram. 


The second result is that the spiral arm model places most of the maser spots in opposite quadrants of the disk, with the blueshifted masers behind the disk midline (farther away from the observer) and the near-systemic and redshifted masers before the midline (closer to the observer). This pattern resembles ionization cones seen in the NLRs of Seyfert galaxies, which result from selective obscuration \citep[e.g.,][]{1994AJ....107.1227W}. In the case of the maser disk, it seems more likely that the asymmetry results from a warped disk or outflows that might elevate molecular clouds out of the disk plane and expose them to the central engine. Alternatively, the central engine could produce an intrinsically anisotropic, polar radiation field \citep[cf.][]{2015ARA&A..53..365N}, and the masers occur in regions that view the central accretion disk more nearly pole-on.

\begin{figure*}[tbh]
  \centering
\includegraphics[width=\textwidth]{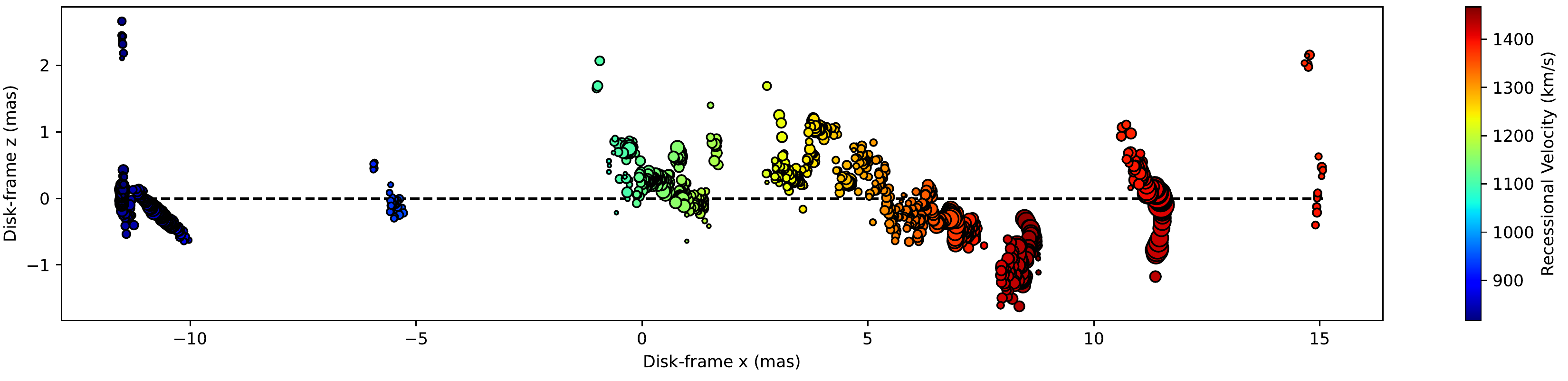}
\caption{The edge-on view of the {deprojected} spiral arm model. The {disk-vertical} ($z$) coordinates were estimated assuming that the minor axis residuals of the model fit result solely from the vertical structure. {The disk-frame $x$ axis coordinate is in the plane of the disk and aligns with the projected major axis on the sky.}\label{fig:spiralArmVertical}}
\end{figure*}

To look for evidence of a warped disk, we deprojected the disk assuming that minor-axis residuals on the sky result only from the vertical structure of the disk. In this case, the disk frame $z$-coordinate of a maser spot is given by
\begin{equation}\label{eq:zcomponent}
    z = r \left(\frac{\sin{\phi}}{\tan{i}} \right) + (X - X_0)\cos{\Omega} - (Y - Y_0)\sin{\Omega},
\end{equation}
where $r$ and $\phi$ are the polar coordinates of the model maser spot. The resulting vertical structure of the disk is shown in Fig.~\ref{fig:spiralArmVertical}. We find that, by design, most of the maser spots lie close to $z = 0$, but there are notable deviations over the G1--R4 region (disk frame $x$ between toughly 0 and 10~mas). One possibility is that these deviations result from an unaccounted for warp in the molecular accretion disk. However, the displacement of the G1--R4 masers is comparable to the length of the maser substructures. To that point, the filamentary substructures align nearly perpendicular to the best-fit disk plane, suggesting outflow along poloidal magnetic field lines (see Sec.~\ref{sec:magneticFields}). The morphology suggests that the asymmetry of the \water{} maser spot distribution does not result from warps or vertical structure in the molecular disk; rather, it seems more likely that the molecular disk is anisotropically heated.

\subsection{An Anisotropic Heating Model for the Molecular Accretion Disk}\label{sec:anisotropicHeating}

\begin{figure*}[tbh]
  \centering
\includegraphics[width=\textwidth]{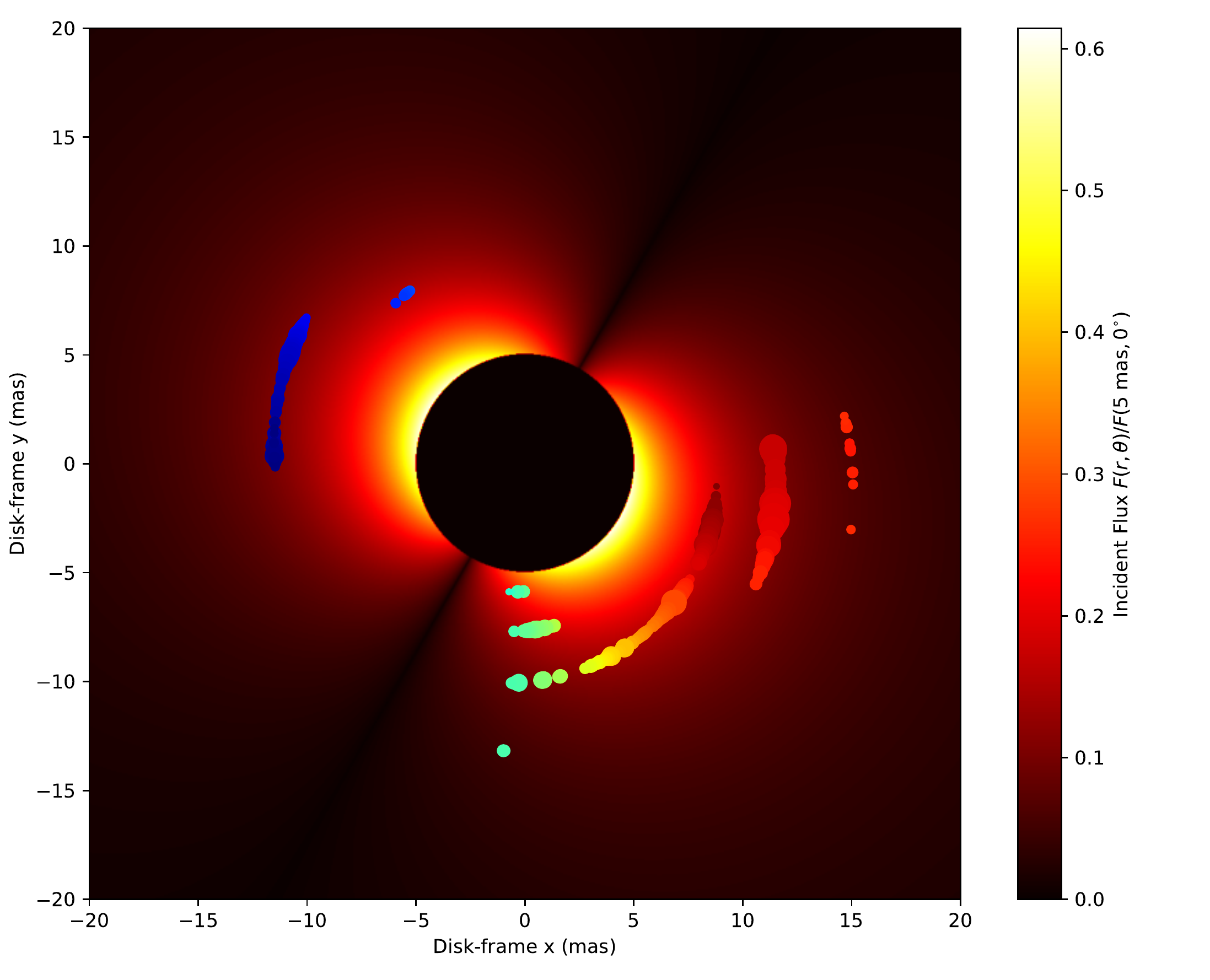}
\caption{The illumination pattern of the central accretion disk on the surrounding molecular accretion disk in the anisotropic heating model. In this model, the central accretion disk has inclination $60\degr{}$, and the polar axis projects onto the sky along PA 11\degr{}. The color scale indicates the flux received by a cloud in the molecular accretion disk; fluxes are normalized to the flux received by a cloud located $r = 5$~mas (0.35~pc) above the accretion disk (i.e. along the polar axis of the accretion disk). The central 5~mas has been masked for the purpose of illustration. The deprojected maser spot positions are plotted as in Fig.~\ref{fig:spiralArmDeprojected}. N.B. this illustration does not include radiative transfer effects within the disk. \label{fig:illuminationPattern}}
\end{figure*}
{So far, \water{} megamaser disks are found mainly in active galactic nuclei \citep[cf.][]{2018ApJ...860..169K}, and the central engine likely plays a role in powering \water{} megamaser emission. Toward understanding the origin of \water{} masers from circumnuclear disks, Neufeld \& Maloney modeled the effects of X-ray heating on molecular gas \citep{1994ApJ...436L.127N,1995ApJ...447L..17N}; they demonstrated that X-ray heating produces a region of enhanced \water{} abundance with temperatures suited to pump the 22~GHz maser transition through collisions. Turning to observational evidence, \cite{2001ApJ...556..694G} demonstrated that the fluxes of the blueshifted and redshifted disk masers of NGC~1068 are correlated over nearly 15~years of monitoring. They also detected a simultaneous flare of blueshifted and redshifted maser features that lasted less than 84~days; however, the projected distance between the maser spots is $\sim 20$~mas (see Fig.~\ref{fig:pvdiagram}), corresponding to a separation $> 4$~LY. The likeliest explanation is that the masers respond to variations of the (unfortunately hidden) central engine \citep[see also][for a theoretical treatment of \water{} maser reverberation]{2000ApJ...542L..99N}.
}

{Motivated by the X-ray heating model for disk megamasers, we consider an anisotropic heating model to explain the distribution of maser spots in the deprojected spiral arms model (Fig.~\ref{fig:spiralArmDeprojected})}. {Specifically}, we consider the radiation pattern produced by a geometrically thin accretion disk with a Thomson-thick atmosphere \citep{1987MNRAS.225...55N}. A cloud located at distance $R$ from the accretion disk receives flux,
\begin{equation}\label{eq:netzerdisk}
    F(r, \theta) \propto \frac{\cos{\theta}(1 + 2 \cos{\theta})}{3R^2}\, ,
\end{equation}
where $\theta$ is the angle between the polar axis of the accretion disk and the radius vector to the cloud. To constrain the model, we assume that the small-scale radio jet along PA $\sim 11\degr{}$ marks the projection of the accretion disk axis on the sky. We further assume that the maser spots occupy quadrants of the molecular disk that receive more AGN continuum flux than the adjacent quadrants. We calculated the illumination of the molecular disk over a grid of accretion disk inclinations and searched for illumination patterns that preferentially heat the maser quadrants of the molecular disk. From inspection of the resulting model grids, the best matches result for accretion disk inclinations between $\approx 56\degr{}$ to $66\degr{}$ (i.e., the northern radio jet axis tilts toward the observer $\approx 24\degr{}$ to $34\degr{}$ from the plane of the sky). Figure~\ref{fig:illuminationPattern} shows the illumination pattern for accretion disk inclination $60\degr{}$. Note that, for this illustration, we have not included the effects of radiative transfer through the molecular disk; rather, our goal was to illustrate the disk heating pattern that results when the larger molecular accretion disk does not align with the inner accretion disk.

Based on the maser data alone, it is unclear which of the following has a greater effect on the distribution of maser spots in the molecular accretion disk: anisotropic heating or the vertical structure of the disk. However, it is unclear why vertical structures preferentially occur in opposite quadrants of the molecular disk unless the disk is anisotropically heated. We revisit to the anisotropic heating model in Sec.~\ref{sec:infrared}. 


\subsection{Constraints on the Mass of the Molecular Accretion Disk}

The presence of spiral arms implies self-gravitation in the disk, which, in turn, can be used to estimate the mass of the molecular disk (see \citealt{1995ApJ...455L.131M} for a similar analysis of NGC~4258) The Toomre $Q$ parameter evaluates the stability of a differentially rotating disk against fragmentation and collapse \citep{1964ApJ...139.1217T}. For a Keplerian disk in orbit around a black hole, 
\begin{equation}
    Q \approx 2 \frac{M_{bh}}{M_{d}} \frac{c}{v_{\phi}}\,,
\end{equation}
where $M_{bh}$ is the central black hole mass, $M_d$ is the disk mass, and $c$ is the characteristic wave speed in the gas, whether the sound speed $c_s \approx 0.08\sqrt{T}$~\kms{} (cf. \citealt{2016ARA&A..54..271K}), or, in a magnetized disk, the Alfv\'{e}n speed, $v_A$ \citep{Kim_2001}. Spiral structure appears when $1 \la Q \la 2$. Normalizing to values appropriate for the disk masers, we find
{
\begin{equation}\label{eqn:diskmass}
    M_d \approx 110 \times 10^3\,\Mo{}\, \left(\frac{M_{bh}}{17\times 10^6\,\Mo{}}\right) 
    \left(\frac{Q}{1.5}\right)^{-1}
    \left(\frac{v_c}{2.6\,\kms{}}\right)
    \left(\frac{v_{\phi}}{300\,\kms{}}\right)^{-1}\,.
\end{equation}
Here, we have scaled the characteristic speed, $v_c$, to the best-fit error floor (Table~\ref{tab:spiralArm}), which is close to the sound speed in warm molecular clouds.
}\explain{We rescaled the velocity based on the error floor rather than the earlier estimate of the Alfven speed. The Alfven speed estimate depended on later hydrostatic equilibrium arguments, which we have since removed.}

We find that the disk mass inferred from the stability arguments is {$\la 1$\%} of the central mass, and the effect on the rotation curve should be small. The exact effect on the rotation curve depends on the radial profile of the surface mass density, $\Sigma(r)$, which is unknown. Assuming a Mestel disk profile for simplicity, $\Sigma(r) \propto r^{-1}$ \citep{1963MNRAS.126..553M}, the characteristic circular speed due only to disk self-gravitation is {$\la 20$~\kms{}}. Combined in quadrature with a characteristic rotation speed $v_{\phi} = 300$~\kms{}, the effect on the rotation curve is {$\la 0.2$\%}. \explain{Minor changes to account for the rescaling of Eqn.~\ref{eqn:diskmass}.}

Based on this estimate of disk mass (Eq.~\ref{eqn:diskmass}) and the geometry of the disk defined by the \water{} maser spots, the mean gas density within the molecular disk is {$\bar{\rho} \approx 7\times 10^{-14}\,\mbox{kg\,m}^{-3}$ ($\bar{n}(\mbox{H}_2) \approx 9\times10^{7}\,\mbox{cm}^{-3}$}). For comparison, \water{} masers trace molecular gas with density {$n(\mbox{H}_2) \approx 10^8$ -- $10^{11}\,\mbox{cm}^{-3}$ \citep{1987ApJ...323..346K,1991ApJ...373..525K,1994ApJ...436L.127N, 1995PNAS...9211427M,2000ApJ...542L..99N}, or}
{$\rho \approx 3\times 10^{-13}\mbox{\ to\  }3 \times 10^{-10}\,\mbox{kg\,m}^{-3}$}. The inferred density contrast between the arm and {the disk average} is {$\rho_A/\rho_d \ga 4$}, {comfortably} within the range of contrasts produced in swing amplification studies \citep{1981seng.proc..111T,1995ApJ...455L.131M}. Based on this simplified stability analysis, the density enhancement produced by the spiral arms may have been necessary to generate \water{} masers in the molecular accretion disk of NGC~1068. \explain{Some benign scaling and language changes to accommodate the change in the velocity scaling of Eqn.~\ref{eqn:diskmass}. It occurred to us that the minimum density for maser emission rather than the critical density is more relevant for the scaling argument. } 

\section{Comparison with Infrared Continuum Images}\label{sec:infrared}

\begin{figure*}[tbh]
  \centering
\includegraphics[width=\textwidth]{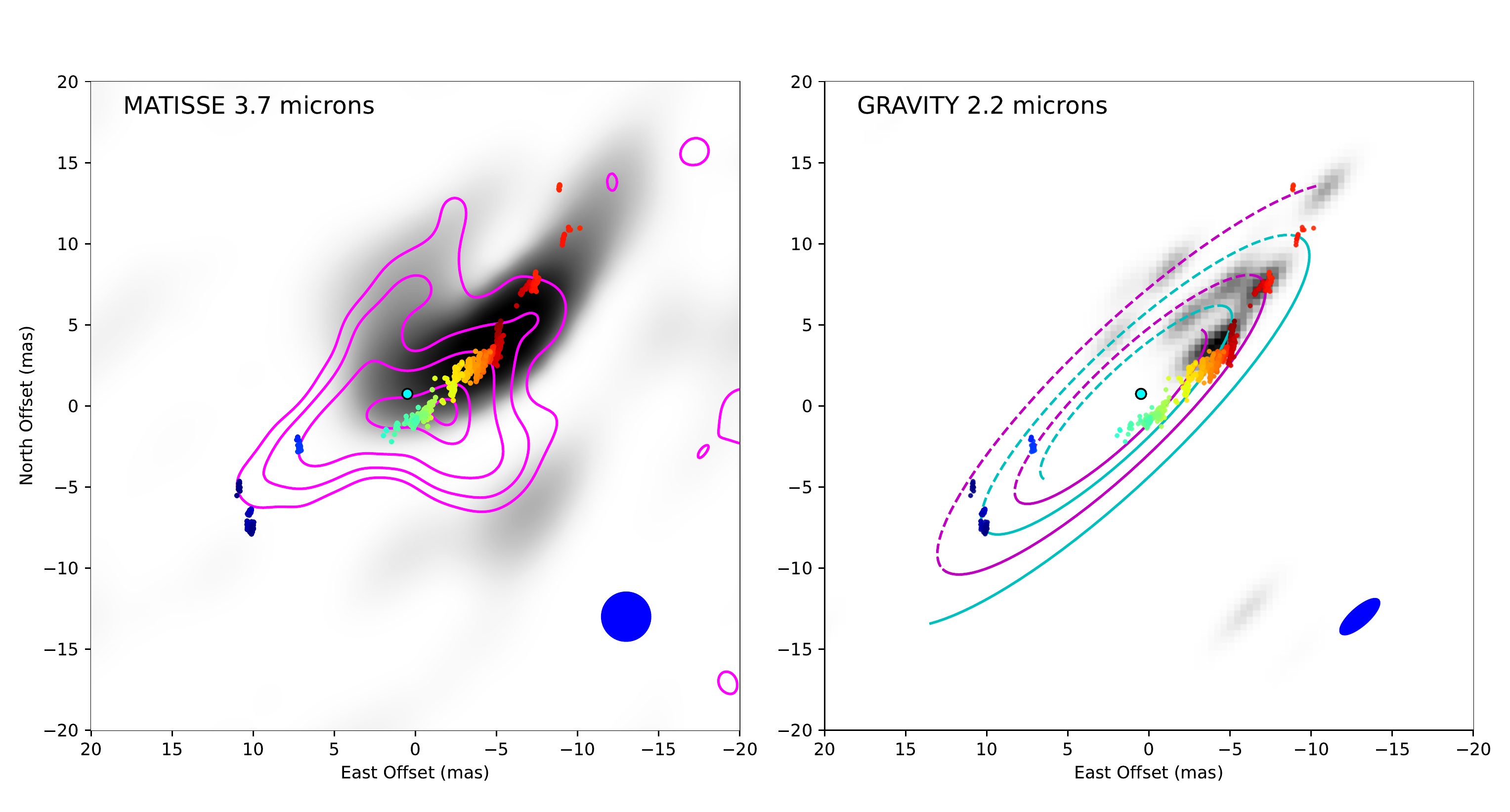}
\caption{Comparison of HSA nuclear continuum and maser positions with VLTI infrared images.  In both panels, the maser spots are plotted as filled dots color-coded by recessional velocity as in Fig.~\ref{fig:s1-skyplot}. The infrared beams are shown as blue-filled ellipses in the lower right corner. {\em Left panel:} In reverse grayscale, the MATISSE 3.7\,\mic{} image from \cite{2022Natur.602..403G}. The 22~GHz continuum is plotted as magenta contours; the contour levels are identical to Fig.~\ref{fig:taperedContinuum}. {\em Right panel:} In inverse grayscale, the GRAVITY 2.2\,\mic{} image from \cite{2020A&A...634A...1G}. The spiral arms model for the maser kinematics is plotted as magenta and cyan curves. The astrometry of the infrared images is based on a cross-correlation between the ALMA 256~GHz continuum image \citep{2019ApJ...884L..28I} and the MATISSE 12\,\mic{} continuum image \citep{2022Natur.602..403G}. Since the masers and 22~GHz continuum are produced from the same data, the relative astrometry is arbitrarily precise. \label{fig:vlti}}
\end{figure*}

The nucleus of NGC~1068 has been observed at roughly mas resolution in the near-infrared and mid-infrared with the Very Large Telescope Interferometer (VLTI): $2.2\,\mic{}$ continuum with the GRAVITY instrument \citep{2020A&A...634A...1G}, and $3.7$--$12\,\mic{}$ with the MATISSE instrument \citep{2022Natur.602..403G}. The infrared continuum images are provided in Fig.~\ref{fig:vlti}. The absolute astrometric calibration for these images is imprecise, so the comparison with radio and mm-wave data has relied on interpretation. The GRAVITY image shows a handful of compact sources that roughly align with the orientation of the disk masers. Brighter sources appear to form a partial ring with radius 3.5~mas (0.25~pc). In their preferred interpretation, \cite{2020A&A...634A...1G} model near-infrared emission as arising from hot, dusty clouds surrounding the central engine and located near the dust sublimation radius. In that registration, the hot dust and masers are coplanar, and the hot dust traces the inner edge of the molecular accretion disk.

Turning to the MATISSE data, \cite{2022Natur.602..403G} derived astrometry based on a cross-correlation between the $12\,\mic{}$ continuum image and the ALMA 256~GHz continuum image of \cite{2019ApJ...884L..28I}. The astrometric precision is about 3~mas across all of the MATISSE bands. Putting some confidence in this alignment, \cite{2022Natur.602..403G} found a morphological agreement between the brightest emission on the MATISSE images and resolved northern and northeastern extensions on the 22~GHz continuum image. Given the proximity in wavelength, the compact sources on the $2.2\mic{}$ GRAVITY image likely associate with the brightest emission on the $3.7\mic{}$ MATISSE image; the GRAVITY image shown in Fig.~\ref{fig:vlti} was registered based on this assumption. To place the positions of the maser spots measured by GG97, \cite{2022Natur.602..403G} assumed that the 22~GHz continuum peak marks the kinematic center of the maser disk. In their registration, the masers fall along a dark lane in the $3.7\,\mic{}$ image, and the brighter infrared emission is located mainly north of the maser disk. Their interpretation is that the mid-infrared continuum traces warm dust in the molecular outflow.

\begin{figure*}[tbh]
  \centering
\includegraphics[width=\textwidth]{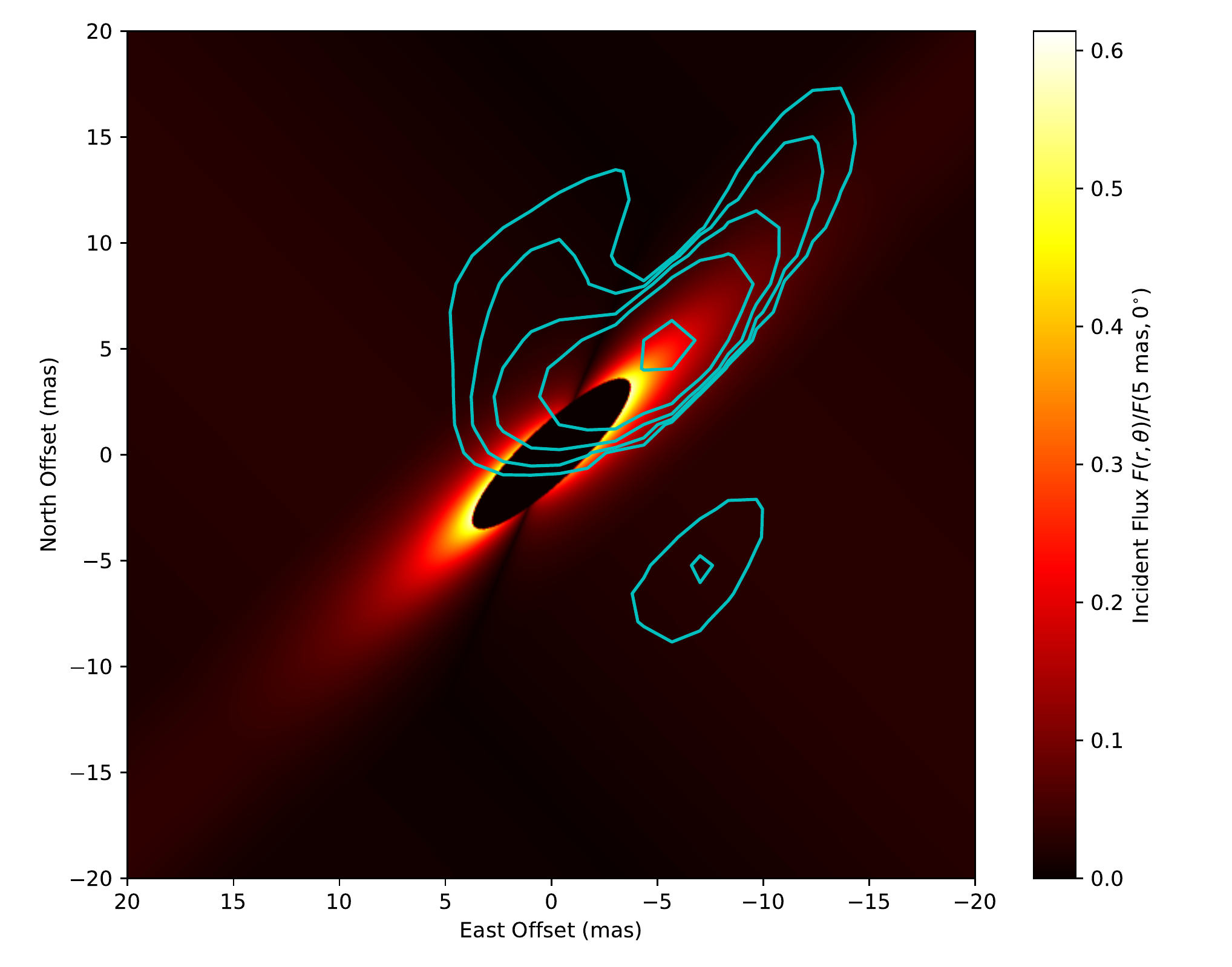}
\caption{The illumination pattern of the central accretion disk on the surrounding molecular accretion disk in the anisotropic heating model as in Fig.~\ref{fig:illuminationPattern} but projected onto the sky. The MATISSE 3.7\,\mic{} continuum image is plotted as cyan contours; the contour levels are 0.07, 0.14, 0.26, 0.51, and 0.97 of the infrared continuum peak. \label{fig:infraredIllumination}}
\end{figure*}

{Because they derive from the same data, our new observations provide a precise registration of the maser spots relative to the 22~GHz continuum (Fig.~\ref{fig:taperedContinuum}).}
{We find that, counter to the registration of \cite{2022Natur.602..403G}, the radio continuum peak is actually located southwest of the kinematic center of the maser disk (Sec.~\ref{sec:astrometry}; Fig.~\ref{fig:taperedContinuum}). As shown in Fig.~\ref{fig:vlti}, the corrected placement of the \water{} maser spots more closely aligns the \water{} maser disk with the GRAVITY near-infrared sources as originally proposed by \cite{2020A&A...634A...1G} but shifts the infrared sources to a larger distance from the kinematic center.}
\explain{Changes to clarify what is changing with the new observations, namely, that \cite{2022Natur.602..403G} placed the masers in the wrong place relative to the 22 GHz continuum. We are exploring the consequences of the correction.}

The inferred dust temperatures are consistent with this registration. Based on modeling of the GRAVITY-MATISSE infrared SEDs, the dust temperatures of the brightest infrared sources are $\sim 700$~K, well below the sublimation temperature for graphite ($T_{sub} \approx 1800$~K) and silicate ($T_{sub} \approx 1500$~K) grains \citep{2022Natur.602..403G}. Scaling from the sublimation radii estimates of \cite{2015ARA&A..53..365N}, the expected grain temperature is
\begin{equation}
T_{gr} \approx 1000\,\left(\frac{L_{AGN}}{10^{45}\,\mbox{ergs~s}^{-1}}\right)^{(1/5.2)} \left(\frac{f(\theta) \times R}{0.8\,\mbox{pc}}\right)^{(1/2.6)}\,\mbox{K,}
\end{equation}
where $L_{AGN}$ is the (unknown) luminosity of the AGN, $R$ is the distance between the AGN and the dusty cloud, $\theta$ is the polar angle between the accretion disk axis and the dusty cloud, and $f(\theta)$ is a correction for the anisotropy of the AGN radiation field. For an isotropic source, $f(\theta) = 1$, and for a thin accretion disk with a Thomson-thick atmosphere, $f(\theta) = \left[\cos{\theta}(1 + 2\cos{\theta})/3\right]^{1/2}$ (cf. Eqn.~\ref{eq:netzerdisk}). From the anisotropic heating model presented in Sec.~\ref{sec:anisotropicHeating}, $\theta \approx 60\degr{}$ in this region of the molecular accretion disk, and $f(\theta) \approx 0.58$. The predicted dust temperature at $R = 0.8$~pc is,
\begin{equation}
T_{gr} \approx 800\,\left(\frac{L_{AGN}}{10^{45}\,\mbox{ergs~s}^{-1}}\right)^{(1/5.2)} \left(\frac{f(\theta)}{0.58}\right)^{(1/2.6)}\left(\frac{R}{0.8\,\mbox{pc}}\right)^{(1/2.6)}\,\mbox{K.}
\end{equation}
Dust temperatures $T_{gr} = 700$~K result for $L_{AGN} \approx 5\times 10^{44}\,\mbox{ergs~s}^{-1}$, comfortably within the wide range of estimates for NGC~1068 (see the discussion in \citealt{2020A&A...634A...1G}). We caution that this estimate carries many assumptions and is not intended to be a precise measure of the (hidden) AGN luminosity. Rather, the result shows that the dust temperatures associated with the brightest VLTI sources are consistent with $R \approx 0.8$~pc and the astrometric registration shown in Fig.~\ref{fig:vlti}. We conclude that the brightest infrared sources on VLTI images trace warm dust in the molecular accretion disk at roughly the orbital radii of the brightest \water{} maser sources. Fainter infrared continuum sources are found north and south of the molecular disk and are probably associated with nuclear outflow \citep[cf.][]{2016ApJ...829L...7G}.

{
Extinction poses a challenge for the \cite{2022Natur.602..403G} registration. To promote \water{} maser emission, the mean molecular gas density is $n(\mbox{H}_2) \ga 10^8\,\mbox{cm}^{-3}$ \citep{1987ApJ...323..346K,1991ApJ...373..525K,1994ApJ...436L.127N, 1995PNAS...9211427M,2000ApJ...542L..99N}. For a disk with 1~mas (0.07~pc) scale-height (Fig.~\ref{fig:spiralArmVertical}) and inclination $\sim 76\degr{}$ (Table~\ref{tab:spiralArm}), the characteristic path-length through the disk to the midplane is about 0.3~pc. The inferred column density to the disk midplane is therefore $N(\mbox{H}_2) \ga 8\times 10^{25}\,\mbox{cm}^{-2}$, corresponding to a K-band extinction of about 9~mag \citep[][and references therein]{1989ApJ...345..245C,10.1111/j.1365-2966.2009.15598.x}. As a result, we should not be able to see near-infrared continuum directly associated with \water{} masers. We can see two ways to reconcile this issue. First, the extinction might be patchy, and we see near-infrared continuum leaking through gaps in maser clouds. Alternatively, the proposed alignment needs to be adjusted by a few mas, perhaps placing the infrared continuum north or northeast of the maser disk, in which case the infrared continuum might trace extraplanar dust associated with the molecular outflow.
}
\explain{The referee made a really good case about extinction being a problem for the Gamez Rosas registration. We raise the issue here and propose two ways to reconcile the observations.}

For the registration of the 22~GHz and infrared data presented here, there is a remarkable asymmetry between regions northwest and southeast of the kinematic center: the infrared continuum is brighter to the northwest, as is the \water{} maser emission. At least part of the asymmetry might be caused by anisotropic heating, as discussed in Sec.~\ref{sec:anisotropicHeating}. To illustrate, Fig.~\ref{fig:infraredIllumination} shows the model illumination pattern of Fig.~\ref{fig:illuminationPattern} projected onto the sky with the inclination and position angle of the \water{} maser disk (Table~\ref{tab:spiralArm}). The 3.7\,\mic{} continuum peak appears to be associated with the more illuminated region of the molecular disk at the northwest. A fainter extension to the northeast and a faint, isolated source to the southwest are more closely aligned with the molecular disk axis and perhaps trace dust in the molecular outflow. However, the anisotropic heating model predicts an infrared continuum peak to the southwest assuming that molecular gas is distributed at least roughly symmetrically in disk azimuth (i.e. allowing for spiral arms or other disk structures). This result supports the argument that the southeast region of the molecular accretion disk is selectively obscured by colder molecular gas in the torus on the pc scale \citep{2022Natur.602..403G}.

\section{Constraints on Magnetic Fields}\label{sec:magneticFields}

The filamentary substructures observed throughout the nuclear \water{} maser disk of NGC~1068, particularly the parallel filaments of the R4 masers, suggest the influence of ordered magnetic fields with size scales comparable to the disk radius (compare with star-forming regions, for example, \citealt{2009ApJ...700.1609M}). We note in passing that the maser filaments qualitatively resemble molecular outflow features in the magnetic accretion disk model proposed by \cite{1992ApJ...385..460E}. This interpretation requires that the molecular gas be partially ionized, as predicted by X-ray heating models for \water{} megamaser emission \citep{1994ApJ...436L.127N,1995ApJ...447L..17N}. The nearly parallel filaments of the R4 region suggest that there are ordered magnetic fields threading the molecular disk and spanning $90\degr{}$ in disk azimuth (see Fig.~\ref{fig:spiralArmDeprojected}). 

Gas motions will tend to stretch and disrupt filaments, but the organization of the R4 filaments hints that an equilibrium between gas motion and magnetic tension has been achieved; for the purpose of order of magnitude estimation, $\rho v_c^2 \approx B^2 / 8\pi$, where $\rho$ is the mass density of the gas, $v_c$ is a characteristic speed of the gas relative to magnetic field line, and $B$ is the characteristic magnetic field strength. If the magnetic field lines are static as viewed by a distant observer, magnetic tension forces will introduce drag on the rotating disk and cause the filaments to curve in the direction of rotation and amplify the azimuthal component of the magnetic field (see, e.g., \citealt{2007A&A...473..701B}). Since the filaments show no strong curvature (except perhaps among the R3 masers), it seems more likely that the larger-scale magnetic field rotates with the molecular disk. In this case, turbulence introduces relative motion between the molecular gas and the large-scale magnetic field lines. The best-fit error floor, $\delta V = 2.6$~\kms{} (Table~\ref{tab:spiralArm}), provides an estimate of turbulent motion in the molecular disk. For equilibrium between turbulent motions and magnetic tension, 
\begin{equation}
    B_{ls} \approx 1.6~\mbox{mG}\,\left( \frac{v_c}{2.6\,\kms{}} \right) \left(\frac{x_e}{10^{-5}}\right)^{1/2} \left(\frac{\rho}{\rho_{max}} \right)^{1/2}\,
\end{equation}
{where $B_{ls}$ refers to the characteristic magnetic field strength of the large-scale magnetic field, $x_e$ is the ionization fraction of the molecular gas \citep[cf.][]{1987ApJ...323..346K, 1994ApJ...436L.127N} and maser emission is quenched at densities exceeding $\rho_{max} = 3.2 \times 10^{-10}\,\mbox{kg\,m}^{-3}$ \citep{1987ApJ...323..346K,1991ApJ...373..525K,1994ApJ...436L.127N, 1995PNAS...9211427M}}. \explain{This expression was updated to take into account the ionization fraction and a corrected estimate of the critical density for water maser emission.} {For comparison, based on infrared polarimetry, \cite{2015MNRAS.452.1902L} find that the hottest dust grains are embedded in a magnetic field of strength $B_{ls} \ga 4~\mbox{mG}$. The magnetic field strength drops below $B_{ls} \sim 1~\mbox{mG}$ outside $r = 3$~pc \citep[from ALMA continuum polarimetry;][]{2020ApJ...893...33L}. We conclude that the strength of the magnetic fields on sub-pc scales is likely sufficient to stabilize the field lines against turbulent motions of a few~\kms{}. Interestingly, it appears that the orientation of the projected magnetic field lines rotates from nearly poloidal to toroidal: PA $\sim 0\degr{}$ in the maser disk (Fig.~\ref{fig:spokesR4}) to PA$\sim 105\degr{}$ in the outer obscuring ``torus'' \citep{2015MNRAS.452.1902L,2020ApJ...893...33L}. }

Zeeman-induced polarization provide {an independent} check on {this} magnetic field estimate, {but the expected polarization is weak}. For \water{} masers, the maximum fractional circular polarization produced by Zeeman-induced hyperfine transitions is, 
\begin{equation}
    \frac{|V_{max}|}{I} = \frac{\mathcal{A} B_{||}}{\Delta v},
\end{equation}
where $|V_{max}|$ is the absolute Stokes-V peak flux density, $I$ is the Stokes-I peak flux density, $\mathcal{A} \approx 0.02$~\kms{}~gauss\mone{} for this transition, $\Delta v$ is the Stokes-I linewidth, and $B_{||}$ is the strength of the magnetic field component parallel to the sight line \citep{1989A&A...214..333F,1992ApJ...384..185N}. For the field strengths and turbulent velocities discussed above, {$|V|/I \ga 0.0012$\%}. \explain{Revised scaling based on corrected magnetic field strength.}

The brightest single maser spot in our data has $I = 260$~mJy, so the prediction is {$|V| \ga 3~\mu\mbox{Jy}$}. \explain{Revised scaling based on corrected magnetic field strength.} Unfortunately, the predicted Stokes-V falls below the detection limit on the channel maps, roughly 5~mJy\,beam\mone{} (see Sec.~\ref{sec:spotMeasurements}). Accepting that our predictions are only rough estimates for the characteristic magnetic field strengths in the molecular disk, we produced Stokes-V maps for the redshifted channels, but no significant signal was detected. Formally, the non-detection places an upper limit on the magnetic field strength, $B_{||} \la 2.5$~gauss. 

Unfortunately, {even if the magnetic fields are much greater than estimated here,} future searches for Zeeman-induced circular polarization in NGC~1068 will likely be limited by confusion. The maser spots are crowded in the sky and in recessional velocity (Figs.~\ref{fig:s1-skyplot} and \ref{fig:pvdiagram}). As a result, the integrated Stokes-V signal may be suppressed by overlapping and oppositely polarized Zeeman features in the spectrum. Unlike the maser disk of NGC~4258 \citep{1998ApJ...508..243H}, NGC~1068 has no strong, isolated \water{} maser features to search for Zeeman-induced circular polarization. On the other hand, linear polarization might probe at least the orientation of the magnetic fields. In particular, for propagation angles $\sim 90\degr{}$, linear polarization fractions approaching 50\% or greater are expected \citep{2019A&A...628A..14L}. Unfortunately, 
our current data do not include linear polarization information, but follow-up observations with full polarization may provide the best constraints on the magnetic field properties of the molecular accretion disk of NGC~1068.

\section{The Kinematics of the Jet Masers}\label{sec:jetmasers}

\begin{figure*}[tbh]
  \centering
\plotone{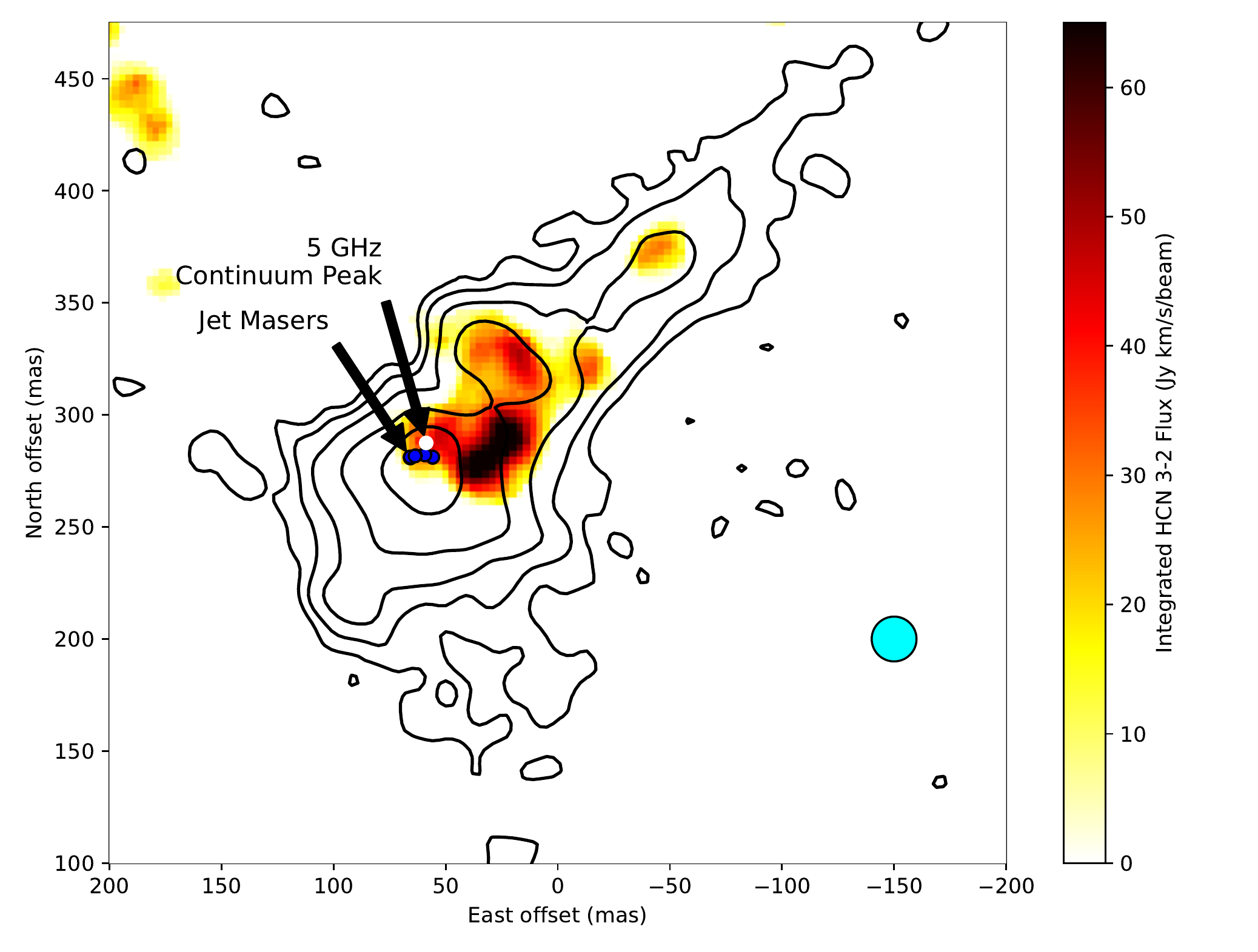}
\caption{The location of the jet masers (blue dots) relative to the peak of the VLBA 5~GHz continuum (white dot), the ALMA 256~GHz continuum image (contours), and integrated HCN~($J=3\rightarrow 2$) emission (colorscale). The ALMA restoring beam is shown as the cyan ellipse on the lower right. The contour levels are 0.019, 0.037, 0.074, 0.15, and 0.30~mJy~beam\mone{}.  \label{fig:HCNoverlay}}
\end{figure*}

The jet masers at continuum component C are blueshifted relative to the systemic velocity of the host galaxy. The surprising result is that they are displaced about 5~mas south of the 5~GHz continuum peak (Fig.~\ref{fig:c-skyplot}). In Fig.~\ref{fig:HCNoverlay}, we compare the location of the jet masers with ALMA maps of 256~GHz continuum and the HCN ($J=3\rightarrow 2$) line (ALMA data from \citealt{2019ApJ...884L..28I}). The HCN emission resolves into a few compact sources with the brightest emission centered $\sim 30$ mas (2~pc) west of the jet masers. For the purpose of discussion, we refer to the clumps of HCN line emission collectively as the component C molecular cloud complex.

\begin{figure*}[tbh]
  \centering
\plotone{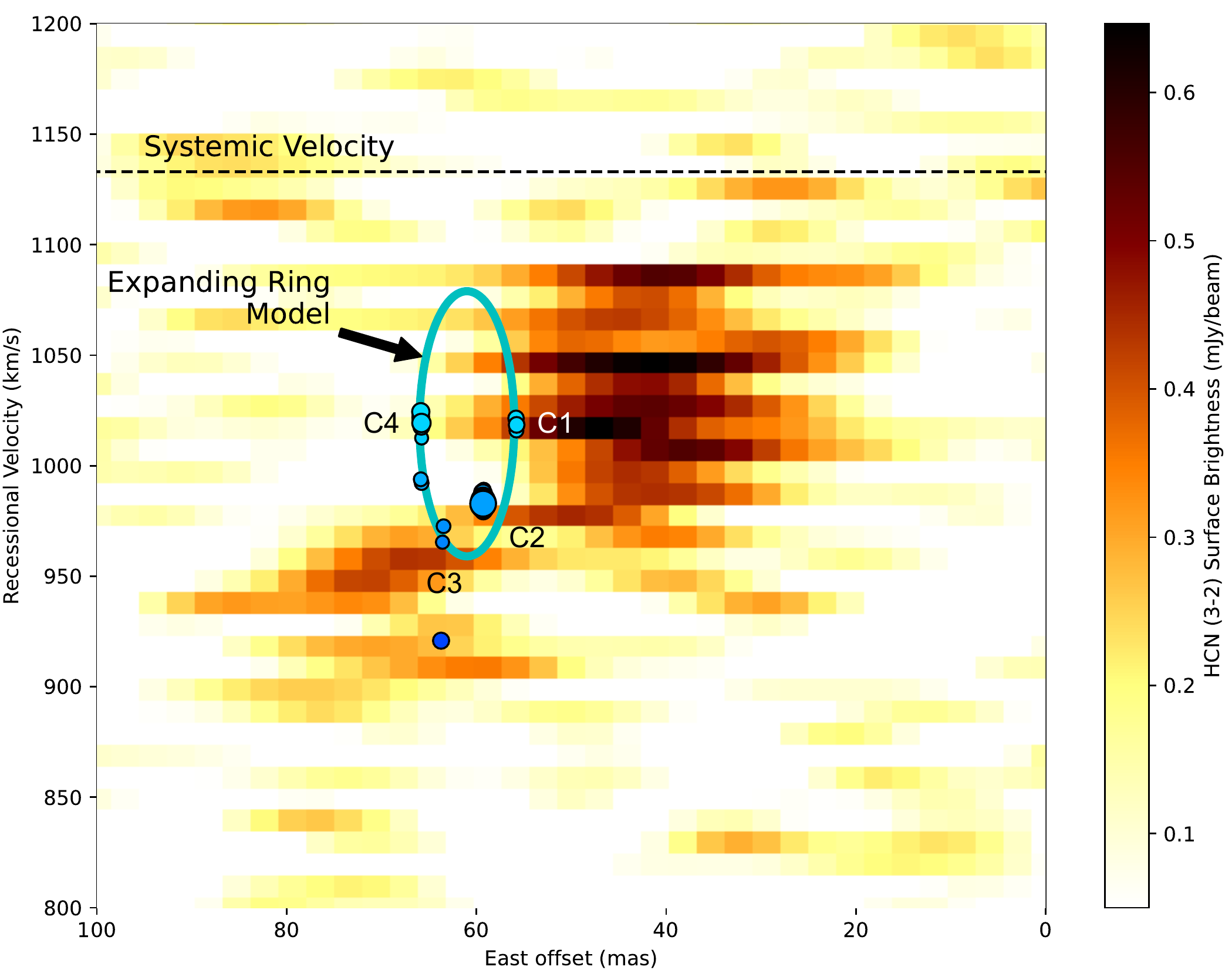}
\caption{The position-velocity diagram of the jet masers (color-filled circles) and HCN~($J=3\rightarrow 2$) emission (colorscale) at component C. The positional slice is east-west through the middle of the jet masers. The spatial resolution of the HCN image is 20~mas. The dashed line indicates the systemic velocity of the host galaxy. The cyan ellipse illustrates the pattern expected for an expanding ring of diameter 10~mas and expansion speed 60~\kms{}. \label{fig:HCNpv}}
\end{figure*}

The east-west \pv{} diagram of the jet masers and molecular cloud complex is provided in Fig.~\ref{fig:HCNpv}. On this diagram, the jet masers show a U-shaped pattern, although the C3 group includes a highly blueshifted spot that overlaps faint HCN emission. The molecular cloud complex spans a similar velocity range but shows additional emission closer to the systemic velocity. No significant HCN emission appears east of the C4 masers. 

Assuming that the masers trace the molecular gas on the near side of some substructure within the molecular cloud complex, the \pv{} diagram is consistent with that expected from an expanding ring viewed nearly edge-on. In this scenario, the C1 and C4 masers move perpendicular to the sight line (that is, C1 moves west and C4 east). Consistent with this picture, the C1 and C4 masers have recessional velocities $\sim 1025$~\kms{}, centered on the mean recessional velocity of the molecular cloud complex. The C2 and C3 masers arise from clumps on the near side of the expanding ring and approach the observer at the speed of expansion, roughly 60~\kms{} relative to the molecular cloud complex.  The C3 masers show the highest relative speeds with recessional velocities reaching 100~\kms{} blueshifted relative to the molecular cloud complex. 

The continuum morphology and the jet maser kinematics might be explained by the impact of the radio jet on the molecular cloud complex in component C (cf. \citealt{1996ApJ...462..740G}). In this scenario, synchrotron continuum emission is enhanced at the resulting shock front; i.e., the 5~GHz continuum peak marks the shock front proper. The post-shock plasma expands, driving a second shock front into the surrounding molecular clouds. Assuming a constant expansion velocity, the kinematic age of the maser ring is $0.35\,\mbox{pc} / 60\,\kms{} \approx 6000\,\mbox{yr}$. As the maser ring expanded to its current diameter, the jet shock front has been advancing roughly northward, away from the central engine. Projected onto the sky, the 5~GHz continuum peak is about 5~mas (0.4~pc) north of the C2 masers. Therefore, assuming that the northward displacement is due to the advancement of the jet shock front over the age of the expanding ring, the average advancement speed is also $\approx 60$~\kms{}.

Relevant to this interpretation, \cite{2017MNRAS.469..994M} proposed that the jet shock powers a secondary wind observed in infrared [\ion{Fe}{2}], [\ion{Si}{6}], and H$_2$ line emission throughout the narrow-line region. The characteristic outflow speeds are $\sim 140$~\kms{} over 100~pc scales, leading to a kinematic age $\approx 7\times 10^5$~years. If these interpretations are correct, the jet masers trace dynamically young molecular clumps that accelerate away from the jet shock and feed the larger-scale and dynamically older outflow. We speculate that the higher-velocity C3 masers may trace accelerated gas located farther from the jet maser ring (i.e. closer to the observer than the expanding ring). It will be interesting to monitor the component C / jet maser region to look for acceleration and the generation of new maser spots as the source evolves.

\section{Conclusions}\label{sec:conclusions}

Based on new HSA observations of the \water{} megamasers of NGC~1068, we have reinterpreted the kinematics of the disk masers and resolved the kinematics of the jet masers. We list the main results and conclusions.
\begin{enumerate}
    \item Using reverse phase calibration on the phase reference source, we obtain astrometric positions of the maser spots and 22~GHz continuum with 0.3~mas precision. By comparing the positions of the maser spots with astrometric images of the 5~GHz continuum, we confirm the close agreement between the morphology of the S1 continuum and the distribution of the disk maser spots. Since it shares the phase reference solutions of the maser spots, the 22~GHz continuum also confirms the positional agreement. 
    \item {Based on the astrometric alignment proposed by \cite{2022Natur.602..403G},} the brightest \water{} masers are {associated with} the brightest infrared sources on VLTI images. It seems likely that these infrared sources trace warm dust in the molecular accretion disk {or its associated outflow}. The dust temperatures are consistent with an AGN luminosity $L_{AGN} \approx 5\times 10^{44}\,\mbox{ergs\,s}^{-1}$.
    \item In contrast to other megamaser disks, the \pv{} diagram of the disk masers shows a peculiar curve between the extrema of the recessional velocities. The masers appear to sample molecular gas in two symmetric spiral arms with pitch angle $\theta_p = 5\degr{}$. 
    \item Based on a deprojection of the spiral arms model, the disk masers are located in an annulus located between roughly 5 and 15~mas (0.35 to 1~pc) from the kinematic center.
    \item The disk masers preferentially fall in opposite quadrants of the molecular accretion disk. We speculate that the masers trace anisotropically heated regions of the molecular accretion disk.
    \item The rotation curve is consistent with Keplerian rotation. The extrapolated rotation curve agrees well with the observed tangential velocity curve of HCN ($J=3\rightarrow 2$), indicating that the rotation curve remains Keplerian out to $r\sim 100$~mas (7~pc). 
    \item The inferred mass inside $r= 5$~mas (0.35~pc) is $17\times10^6~\Mo$.
    \item Based on disk stability arguments, the mass of the molecular disk is {$\approx 110\times 10^3$~\Mo{}}. 
    \item {The velocity error floor of the spiral disk model} is $\delta V = 2.6$~\kms{}, much lower than the previous estimates {of the turbulent velocity}. We note that this value is comparable to the sound speed in warm molecular gas.
    \item On mas scales, the disk masers arrange into linear, filamentary structures, suggesting the influence of pc-scale magnetic fields {with characteristic field strengths $\ga 1.6$~mG}.
    \item The jet masers appear to trace an expanding ring with characteristic expansion velocity $\sim 60~\kms{}$. The kinematic age is about 6000~years.
    \item The radio continuum source C is displaced about 5~mas north of the jet masers. We propose that the continuum source traces an advancing shock as the radio jet penetrates the molecular cloud complex near component C. Assuming that the shock front has advanced northward from the jet masers over 6000~years, its advancement speed is also $\sim 60$~\kms{}. 
\end{enumerate}

\begin{acknowledgments}
The authors thank the participants of the TORUS 2018 workshop (Puerto Varas, Chile) for the helpful comments and conversations that motivated this work. We also thank the anonymous referee for their helpful and constructive review that greatly improved the manuscript. This paper makes use of data obtained from the NSF's VLBA and VLA, operated by the National Radio Astronomy Observatory. The National Radio Astronomy Observatory is a facility of the National Science Foundation operated under cooperative agreement by Associated Universities, Inc. The GBT is part of the Green Bank Observatory, which is a facility of the National Science Foundation operated under cooperative agreement by Associated Universities, Inc.
\end{acknowledgments}

\vspace{5mm}
\facilities{VLBA, VLA, GBT, HSA}
\software{AIPS \citep{1985daa..conf..195W,1990apaa.conf..125G,2003ASSL..285..109G}, astropy \citep{2013A&A...558A..33A}, emcee \citep{2013PASP..125..306F}, DIFMAP \citep{1997ASPC..125...77S}, PyDREAM \citep{10.1093/bioinformatics/btx626}}

\appendix

\section{Subgroup Analysis and Results}\label{app:substructure}

\begin{figure*}[tbh]
  \centering
\plotone{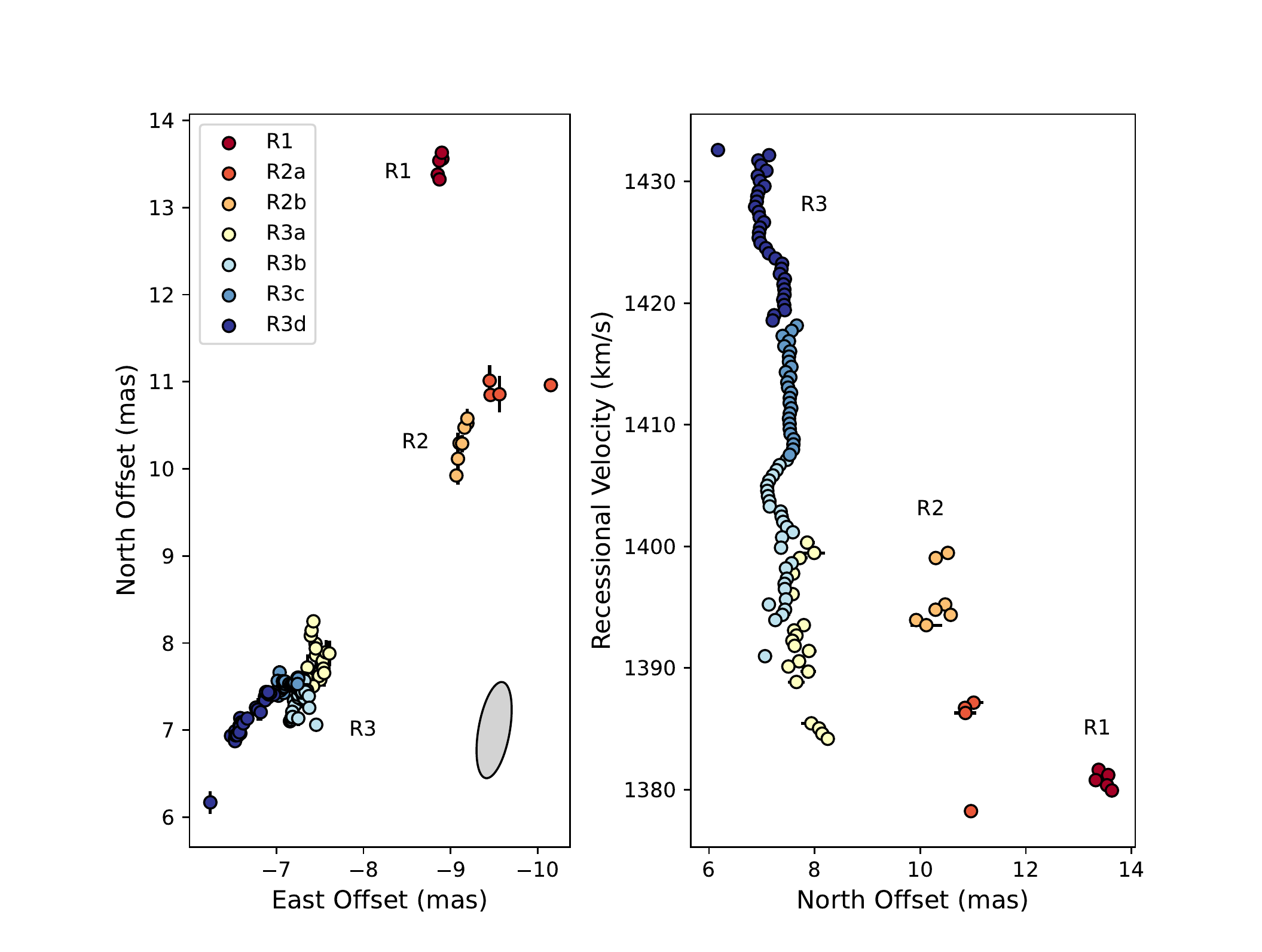}
\caption{Close-up of the R1--3 maser groups, plotted as in Fig.~\ref{fig:spokesR4}.   \label{fig:spokesR1t3}}
\end{figure*}

\begin{figure*}[tbh]
  \centering
\plotone{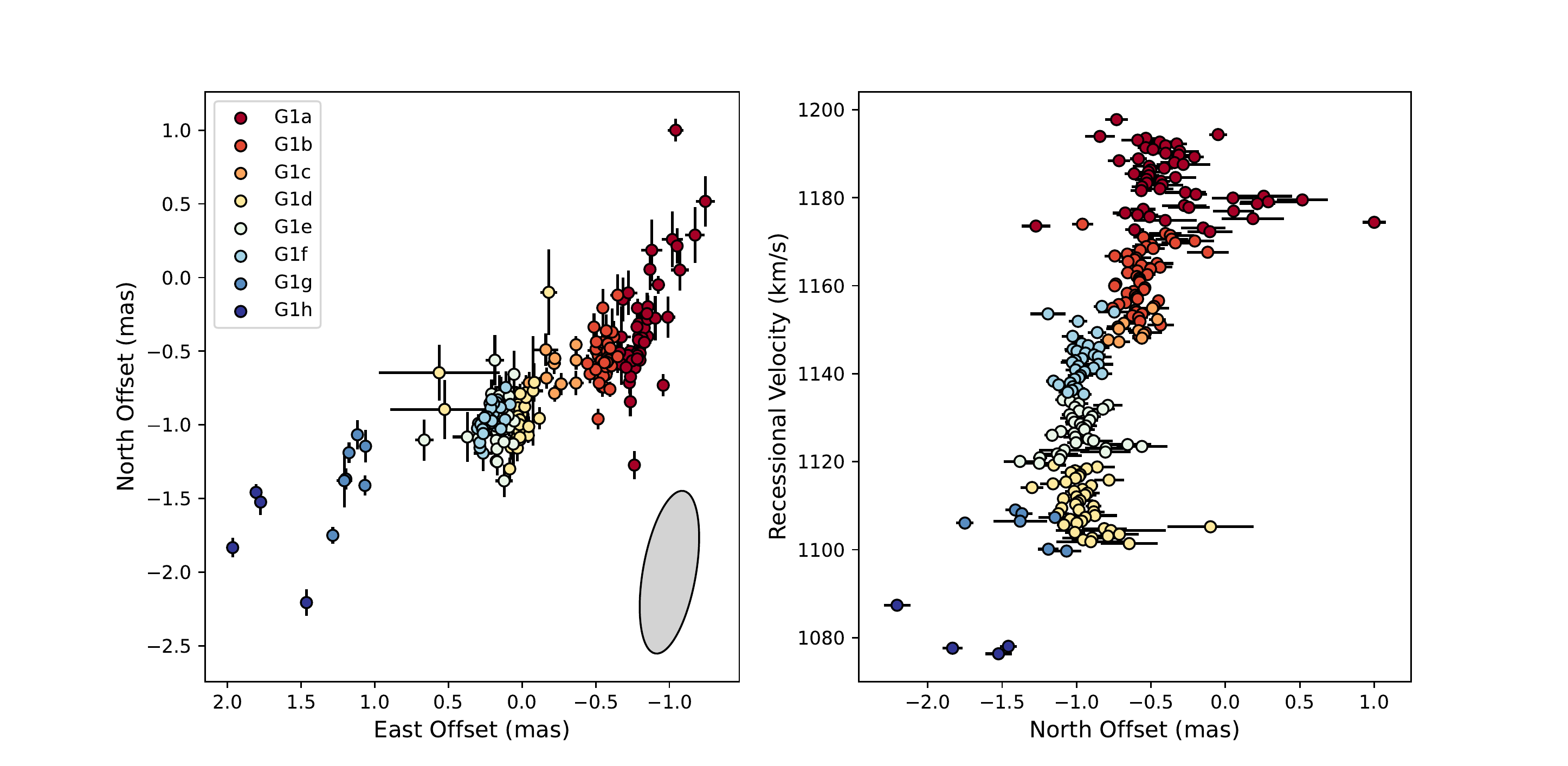}
\caption{Close-up of the G1 maser group, plotted as in Fig.~\ref{fig:spokesR4}.  \label{fig:spokesG1}}
\end{figure*}

\begin{figure*}[tbh]
  \centering
\plotone{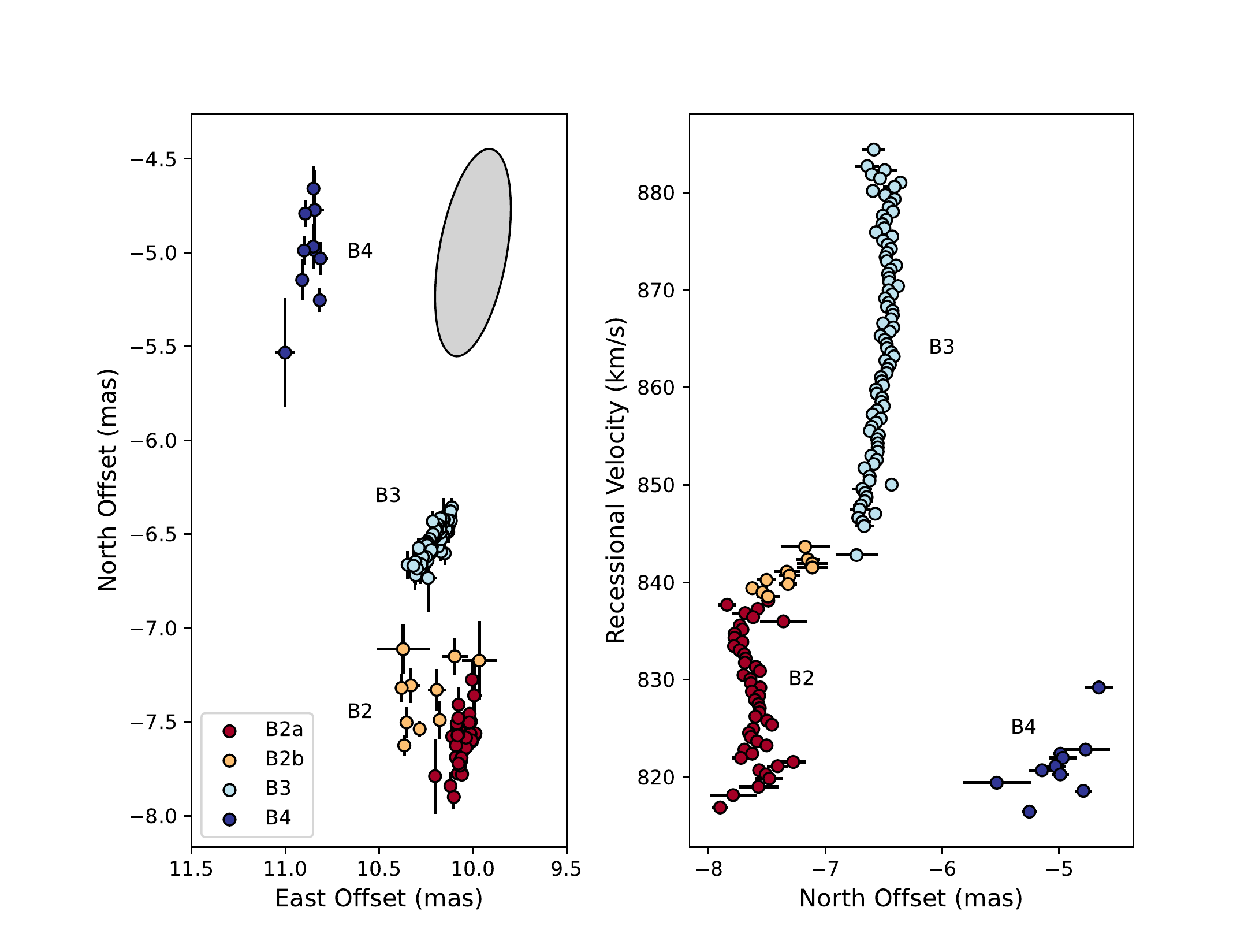}
\caption{Close-up of the B2--4 maser groups, plotted as in Fig.~\ref{fig:spokesR4}. \label{fig:spokesB24}}
\end{figure*}

Close-ups of the sky distribution of the individual disk maser groups are provided in Figures~\ref{fig:spokesR4} and \ref{fig:spokesR1t3} -- \ref{fig:spokesB24}. Declination-velocity diagrams are included to highlight north--south velocity gradients within the maser groups. Group R4, in particular, breaks up into remarkable parallel filaments (Figure~\ref{fig:spokesR4}). 

We used k-means clustering \citep{Forgy1965ClusterAO, macqueen1967classification} to classify individual substructures for analysis. This clustering technique does not determine the optimal number of clusters to assign; instead, we iteratively varied the number of clusters and inspected the cluster assignments by eye. The goal was to find the minimum number of clusters required to separate clear groupings in position and velocity. Using this approach, we indentified 34 subgroups and labeled them by their parent group name and a lowercase suffix (e.g., R2a, R2b, etc.).  No subgroups were found for groups R1, B3, and B4.

We calculated the properties of the subgroups using flux-weighted moments of the sky coordinates, and we performed a linear fit to the major axis offset and recessional velocities to estimate the mean velocity gradient. Table~\ref{tab:substructures} lists the summary properties of the subgroups. Uncertainties were determined using a Monte Carlo method. In each Monte Carlo trial, the spot coordinates and flux densities were drawn from a normal distribution scaled to the measurement uncertainties. The flux-weighted moments and mean velocity gradients were recalculated for each trial. After $10^4$~trials, the uncertainties of the properties of the subgroups were estimated using the standard deviation for all the trials.

\begin{deluxetable*}{lD@{\,$\pm$\,}DD@{$\pm$}DD@{\,$\pm$\,}DD@{\,$\pm$\,}DD@{\,$\pm$\,}DD@{\,$\pm$\,}DD@{\,$\pm$\,}D}
\tablecaption{Maser Subgroup Properties\label{tab:substructures}}
\tablehead{\colhead{Name} & \multicolumn4c{East Offset} & \multicolumn4c{West Offset} & \multicolumn4c{$V_R$} & \multicolumn4c{Major Axis} & \multicolumn4c{Minor Axis} & \multicolumn4c{PA} & \multicolumn4c{$V_R^{\prime}$} \\
\colhead{(1)} & \multicolumn4c{(2)} & \multicolumn4c{(3)} & \multicolumn4c{(4)} & \multicolumn4c{(5)} & \multicolumn4c{(6)} & \multicolumn4c{(7)} & \multicolumn4c{(8)}}
\decimals
\startdata
R1 & -8.88 & 0.02 & 13.46 & 0.07 & 1380.983 & 0.008 & 0.40 & 0.02 & 0.05 & 0.01 & -8 & 1 & -2.8 & 0.4 \\
R2a & -9.63 & 0.03 & 10.9 & 0.1 & 1384.93 & 0.04 & 0.97 & 0.02 & 0.30 & 0.07 & 1 & 9 & -7 & 3 \\
R2b & -9.13 & 0.03 & 10.3 & 0.1 & 1395.678 & 0.003 & 0.8 & 0.1 & 0.09 & 0.04 & -11 & 3 & 4 & 1 \\
R3a & -7.46 & 0.03 & 7.8 & 0.1 & 1392.66 & 0.03 & 0.71 & 0.05 & 0.274 & 0.008 & -2 & 2 & -11 & 2 \\
R3b & -7.24 & 0.01 & 7.29 & 0.04 & 1402.02 & 0.07 & 0.57 & 0.08 & 0.15 & 0.02 & -20 & 3 & -8 & 2 \\
R3c & -7.135 & 0.003 & 7.53 & 0.01 & 1412.04 & 0.02 & 0.248 & 0.004 & 0.12 & 0.01 & 25 & 2 & 13 & 9 \\
R3d & -6.668 & 0.008 & 7.09 & 0.03 & 1425.591 & 0.003 & 0.970 & 0.008 & 0.13 & 0.01 & -36.5 & 0.5 & -10.7 & 0.4 \\
R4a & -5.13 & 0.01 & 4.66 & 0.05 & 1452.04 & 0.01 & 0.90 & 0.06 & 0.241 & 0.006 & -2 & 1 & 14.1 & 0.6 \\
R4b & -5.09 & 0.02 & 3.56 & 0.07 & 1432.6 & 0.1 & 1.22 & 0.05 & 0.27 & 0.02 & -6 & 1 & 8.6 & 0.6 \\
R4c & -4.68 & 0.02 & 3.17 & 0.08 & 1380.88 & 0.07 & 0.85 & 0.04 & 0.100 & 0.008 & -15.5 & 0.9 & 19.1 & 0.6 \\
R4d & -4.512 & 0.008 & 3.13 & 0.04 & 1362.04 & 0.03 & 0.680 & 0.009 & 0.103 & 0.004 & -18.1 & 0.3 & 25.6 & 0.4 \\
R4e & -4.22 & 0.03 & 2.8 & 0.1 & 1339.9 & 0.2 & 1.4 & 0.1 & 0.207 & 0.006 & -7 & 1 & 9.4 & 0.6 \\
R4f & -3.79 & 0.06 & 2.3 & 0.2 & 1312.18 & 0.03 & 1.5 & 0.1 & 0.5 & 0.2 & 0 & 5 & -9.2 & 0.8 \\
R4g & -3.33 & 0.03 & 2.5 & 0.1 & 1292.24 & 0.08 & 1.1 & 0.2 & 0.256 & 0.008 & -1 & 3 & -5 & 2 \\
R4h & -3.19 & 0.02 & 1.92 & 0.07 & 1276.37 & 0.07 & 0.5 & 0.2 & 0.23 & 0.02 & -10 & 20 & 4 & 6 \\
R4i & -2.60 & 0.01 & 2.33 & 0.05 & 1262.47 & 0.02 & 0.40 & 0.03 & 0.20 & 0.02 & -40 & 20 & 20 & 10 \\
R4j & -2.58 & 0.07 & 1.8 & 0.2 & 1251.61 & 0.04 & 0.75 & 0.09 & 0.5 & 0.1 & -20 & 30 & 3 & 4 \\
R4k & -2.34 & 0.03 & 1.0 & 0.1 & 1236.50 & 0.08 & 0.64 & 0.07 & 0.23 & 0.03 & -5 & 4 & 2 & 1 \\
R4l & -1.93 & 0.03 & 1.61 & 0.08 & 1228.782 & 0.001 & 0.48 & 0.02 & 0.11 & 0.04 & 34 & 4 & -1.7 & 0.3 \\
R4m & -1.5 & 0.2 & 0.8 & 0.3 & 1217.490 & 0.002 & 2.5 & 0.1 & 0.2 & 0.2 & 17 & 7 & -0.4 & 0.1 \\
G1a & -0.84 & 0.04 & -0.3 & 0.1 & 1183.42 & 0.03 & 1.17 & 0.08 & 0.263 & 0.008 & -18 & 2 & -5 & 0.5 \\
G1b & -0.56 & 0.02 & -0.57 & 0.05 & 1160.14 & 0.06 & 0.40 & 0.07 & 0.14 & 0.02 & -6 & 5 & 10 & 4 \\
G1c & -0.24 & 0.05 & -0.63 & 0.07 & 1149.87 & 0.02 & 0.415 & 0.008 & 0.29 & 0.01 & -40 & 10 & 8 & 2 \\
G1d & 0.04 & 0.04 & -0.95 & 0.08 & 1110.09 & 0.04 & 0.7 & 0.2 & 0.4 & 0.2 & -10 & 20 & -10 & 3 \\
G1e & 0.18 & 0.03 & -0.98 & 0.08 & 1128.09 & 0.08 & 0.52 & 0.05 & 0.34 & 0.02 & -17 & 6 & 6 & 4 \\
G1f & 0.23 & 0.03 & -0.97 & 0.07 & 1142.38 & 0.07 & 0.34 & 0.02 & 0.14 & 0.01 & -26 & 4 & 20 & 2 \\
G1g & 1.16 & 0.03 & -1.3 & 0.1 & 1105.82 & 0.02 & 0.75 & 0.04 & 0.18 & 0.02 & -16 & 2 & -7.3 & 0.7 \\
G1h & 1.77 & 0.03 & -1.72 & 0.07 & 1079.47 & 0.06 & 1.06 & 0.03 & 0.462 & 0.007 & 23.3 & 0.9 & -12.8 & 0.3 \\
B1a & 7.14 & 0.02 & -2.54 & 0.09 & 939.64 & 0.07 & 0.45 & 0.06 & 0.14 & 0.01 & -2 & 3 & -2 & 4 \\
B1b & 7.27 & 0.03 & -2.0 & 0.1 & 919.457 & 0.002 & 0.28 & 0.03 & 0.067 & 0.005 & -15 & 2 & 1 & 1 \\
B2a & 10.05 & 0.01 & -7.62 & 0.04 & 828.82 & 0.03 & 0.39 & 0.03 & 0.10 & 0.02 & -11 & 3 & -9 & 3 \\
B2b & 10.27 & 0.06 & -7.4 & 0.1 & 840.56 & 0.04 & 0.6 & 0.1 & 0.42 & 0.03 & -20 & 20 & 7 & 1 \\
B3 & 10.195 & 0.009 & -6.51 & 0.03 & 863.938 & 0.004 & 0.31 & 0.02 & 0.08 & 0.01 & -32 & 3 & 80 & 10 \\
B4 & 10.88 & 0.04 & -5.0 & 0.1 & 821.556 & 0.009 & 0.9 & 0.1 & 0.15 & 0.02 & -9 & 1 & 8 & 2 \\
\enddata
\tablecomments{The properties of the maser subgroups were determined from flux-weighted moments of spot coordinates. The columns are (1) the name of the subgroup; (2) east offset centroid (mas); (3) north offset centroid (mas);  (4) recessional velocity centroid (\kms); (5) major axis (mas); (6) minor axis (mas); (7) position angle (\degr); and (8) mean recessional velocity gradient along the major axis (\kms\,mas\mone), where $V_R^{\prime} > 0$ if velocities increase to the north.}
\end{deluxetable*}

\section{Markov Chain Monte Carlo Analysis}\label{sec:mcmcdetails}

We used a forward modeling approach and Markov Chain Monte Carlo (MCMC) analysis to fit kinematic models to the disk maser spot data. For simplicity, the masers are assumed to occupy a flat disk with inclination $i$ and position angle $\Omega$. Given a trial set of orbital parameter values, we calculated the orbit in cylindrical coordinates in the disk frame $(r, \phi, z=0)$ using discrete steps in azimuth, $\delta\phi = 0\fdg{}1$. The models project onto the sky according to,
\begin{equation}
\begin{split}
X & = X_0 + r \left(\sin{\Omega}\cos{\phi} - \cos{\Omega}\cos{i}\sin{\phi} \right),  \\
Y & = Y_0 + r \left(\cos{\Omega}\cos{\phi} + \sin{\Omega}\cos{i}\sin{\phi} \right),\, \mbox{and} \\
V & = \vsys{} + v_r\, \sin{i}\sin{\phi} + v_\phi\,\sin{i}\cos{\phi}, 
\end{split}\label{eqn:skyprojection}
\end{equation}
where, following the notation of \cite{2013ApJ...775...13H}, $X$ and $Y$ are the sky coordinates, measured as offset angles; $V$ is the recessional velocity; $X_0$ and $Y_0$ are the sky coordinates of the kinematic center; \vsys{} is the systemic recessional velocity; and $v_r$ and $v_{\phi}$ are disk frame orbital speeds in the radial and azimuthal directions, respectively. 

After projecting the model to the sky coordinates, each maser spot was matched to the point on the model orbit that minimizes the Cartesian distance, $\sqrt{(\Delta X/S_x)^2 + (\Delta Y/S_y)^2 + (\Delta V/S_V)^2}$, where $S_X$, $S_Y$, and $S_V$ are measurement uncertainties; see \cite{2020NatAs...4.1170C} for a similar approach to modeling maser kinematics. The goodness of fit for a set of model parameters $\boldsymbol{\theta}$ is given by the unnormalized posterior probability,
\begin{equation}\label{eqn:logprob}
\log P(\boldsymbol{\theta} | \boldsymbol{Q}) = -\frac{1}{2}  \sum_j{\left(\frac{(Q_j- \tilde{Q}_j[\boldsymbol{\theta}])^2}{S_j^2} + \log{\left(2\pi S_j^2 \right)} \right)} + \log{P(\boldsymbol{\theta})},
\end{equation}
where $\boldsymbol{Q}$ is the vector of measurements (i.e., sky positions and recessional velocities), $\boldsymbol{Q} = (Q_1, Q_2, \ldots)$; $\tilde{Q}_j$ are the best-matching model coordinates; $S_j$ is the measurement uncertainty of $Q_j$; $\boldsymbol{\theta}$ is the vector of model parameter values; and $P(\boldsymbol{\theta})$ is the {\em prior} probability, i.e., constraints on the model parameters prior to the model fit. In each iteration of the MCMC analysis, the parameters $\boldsymbol{\theta}$ are randomly updated, $\boldsymbol{\theta}_k \rightarrow \boldsymbol{\theta}_{k+1}$, and the updated parameters are accepted or rejected according to the Metropolis criterion \citep{1953JChPh..21.1087M}. 

 The azimuthal grid spacing of the discretized model introduces some small systematic uncertainty. For $\delta\phi = 0\fdg{}1$, the projected model coordinates are spaced by $\sim 0.015$~ mas in the sky and by $\sim 0.5$~\kms{} in recessional velocity. However, this projected gridding in the sky is much finer than the size of individual subgroups (Table~\ref{tab:substructures}), and the velocity gridding is only slightly larger than the channel width but smaller than the velocity range spanned by individual subgroups. Therefore, the intrinsic substructure within the maser groups is likely to have a greater systematic impact on the fit. To this end, we estimate systematic uncertainties by introducing the parameters $\delta X$, $\delta Y$, and $\delta V$, which are added in quadrature to the measurement uncertainties. These terms are called ``error floors'' in the terminology of \cite{2013ApJ...775...13H}. Note that, with the inclusion of these error floors as parameters, the uncertainties $S_j$ in Eq.~\ref{eqn:logprob} depend on the parameter values: $S_j \rightarrow S_j(\boldsymbol{\theta})$.

For the expanding ring and elliptical orbit models, it is clear the R1 -- R3 masers are not part of the common orbit proposed for the R4 -- B4 maser groups. Recessional velocities decrease with offset of the major axis (Fig.~\ref{fig:pvdiagram}), suggesting that the R1--R3 masers trace the molecular gas outside the orbit(s) traced by the R4--B4 masers. For simplicity and consistency with earlier interpretations (GG97, \citealt{Kumar99}, \citealt{2002A&A...395L..21H}, \citealt{LB03}), we model the R1--R3 masers as tracing the disk midline. As the midline of a flat (unwarped) disk would trace a line on the sky, we anticipate a poor fit to the R1 masers, since they do not line up with the R2 and R3 masers (Fig.~\ref{fig:s1-skyplot}). 

Accepting that some maser groups might not trace orbital motion but rather, say, outflow, we used a mixture model to account for potential outliers (see, e.g., \citealt{GEWEKE20073529,2010arXiv1008.4686H}). Using the formalism commonly employed for mixture modeling, the orbit model intended to fit most of the maser spots is the ``foreground'' (fg), and the model for outliers is the ``background'' (bg). As we have no prior knowledge of the kinematic behavior of potential outliers, we model the background as a three-dimensional Gaussian distribution in sky positions and recessional velocity. The use of a non-physical Gaussian background model has proven to be a robust method of identifying and separating outliers from the foreground model in the absence of any prior constraints on the background \citep{1997upa..conf...49P, 2010arXiv1008.4686H}. By itself, the background model introduces six (6) additional parameters: the means of the background coordinates $(X_{bg}, Y_{bg}, V_{bg})$ and their standard deviations $(\delta X_{bg},\,\delta Y_{bg},\,\delta V_{bg})$. The sums in the log posterior probability, Eq.~\ref{eqn:logprob}, now include terms both for the foreground and background models, with the foreground terms weighted by $P_{fg}$, the probability that any given data point belongs to the foreground, and the background terms weighted by $(1 - P_{fg})$. Here, $P_{fg}$ is an additional fitted parameter. Ideally, the model provides a good description of all of the data, so $P_{fg} \approx 1$; however, we placed no prior constraints other than $0 \leq P_{fg} \leq 1$. In addition, the probability that any single maser spot belongs to the foreground model can be estimated {\em post hoc} from the MCMC results (cf. \citealt{2010arXiv1008.4686H}). This estimate was used to identify the likely outliers for each kinematic model.

We used the MCMC code PyDREAM \citep{10.1093/bioinformatics/btx626} to perform the model fitting analysis.  For each model, we run PyDREAM with three parallel chains and $N = 2\times 10^5$ iterations. To evaluate convergence, we calculated the integrated autocorrelation time (IAT) for each chain of parameter values. The IAT is an estimate of the number of MCMC samples between uncorrelated samples in the chain; put another way, $N / {\rm IAT}$ is an estimate of the number of independent samples after accounting for autocorrelation within a chain of parameter values \citep{2018ApJS..236...11H}. We used the function {\tt autocorr.integrated\_time} from the {\em emcee} software package to calculate the IAT \citep{2013PASP..125..306F}. We consider models with $N / {\rm IAT} > 100$ for all parameters to have converged. In practice, the models converged in a few thousand iterations and $N / {\rm IAT} > 1000$ for all parameters.

In Bayesian analysis, the marginal likelihood $P(\boldsymbol{Q})$, sometimes called the {\em evidence}, is used for model selection. The marginal likelihood is the probability of observing the data $\boldsymbol{Q}$ given the model, and the model with the greater marginal likelihood is preferred, as it better represents the data.  We estimate the marginal likelihood using both the widely applicable Bayes information criterion (WBIC), following the convention of \cite{Friel2017InvestigationOT}, and the thermodynamic integration estimator (TIE; \citealt{10.1214/ss/1028905934,10.2307/1390653}). As shown in Table~\ref{tab:wbic}, both estimators consistently and clearly favor the spiral arms model for the disk masers.


\end{document}